\documentclass{emulateapj}

\usepackage{graphicx}
 
\usepackage{color}
\usepackage{rotating}
\shorttitle{Search for bright nearby M dwarfs with Virtual Observatory tools}

\shortauthors{Aberasturi et al.}

\usepackage{natbib}
\usepackage{longtable}
\usepackage{color}

\begin{document}
%% LaTeX will automatically break titles if they run longer than
%% one line. However, you may use \\ to force a line break if
%% you desire.

\title{Search for bright nearby M dwarfs with Virtual Observatory tools}

\author{M. Aberasturi, J.~A. Caballero, B. Montesinos, M.~C. G\'{a}lvez-Ortiz, E. Solano and E.~L. Mart\'{\i}n}
\affil{Centro de Astrobiolog\'{\i}a (CSIC-INTA), 
   Departamento de Astrof\'{\i}sica, 
   PO Box 78, E-28691 Villanueva de la Ca\~{n}ada, Madrid, Spain}

\begin{abstract}
Using Virtual Observatory tools, we cross-matched the Carlsberg Meridian 14 and the 2MASS Point Source catalogs to select candidate nearby bright M dwarfs distributed over $\sim$25,000\,deg$^2$. Here, we present reconnaissance low-resolution optical spectra for 27 candidates that were observed with the Intermediate Dispersion Spectrograph at the 2.5\,m Isaac Newton Telescope ($\mathcal R \approx$ 1600). 
We derived spectral types from a new spectral index, $\Re$, which measures the ratio of fluxes at 7485--7015$\AA$ and 7120--7150$\AA$.
We also used VOSA, a Virtual Observatory tool for spectral energy distribution fitting, to derive effective temperatures and surface gravities for each candidate.
The resulting 27 targets were M dwarfs brighter than $J$ = 10.5\,mag, 16 of which were completely new in the Northern hemisphere and 7 of which were located at less than 15\,pc. 
For all of them, we also measured H$\alpha$ and Na~{\sc i} pseudo-equivalent widths, determined photometric distances, and identified the most active stars.
The targets with the weakest sodium absorption, namely J0422+2439 (with X-ray and strong H$\alpha$ emissions), J0435+2523, and J0439+2333, are new members in the young Taurus-Auriga star-forming region based on proper motion, spatial distribution, and location in the color-magnitude diagram, which reopens the discussion on the deficit of M2--4 Taurus stars.  
Finally, based on proper motion diagrams, we report on a new wide M-dwarf binary system in the field, LSPM J0326+3929EW.
\end{abstract}

\keywords{astronomical databases: miscellaneous --
open clusters and associations: individual (Taurus-Auriga) --
stars: late-type --
stars: low-mass --
stars: pre-main sequence --
virtual observatory tools
}

\section{Introduction}

\label{introduction}

M dwarfs are the most common stars in the universe. Not only is the closest star to the Sun an M dwarf (Proxima Centauri), but also  66\,\% of the nearest stars in our Galactic neighborhood ($d <$ 10 pc) are M dwarfs\footnote{\tt http://www.recons.org/}. 
However, after the first key proper-motion surveys in the first decades of the 20th century (\citealt{1915ApJ....41..187V}; \citealt{1919VeHei...7..195W}; \citealt{1939AJ.....48..163R}), the famous catalogs of \citet{1969VeARI..22....1G}, \citet{1971lpms.book.....G, 1978LowOB...8...89G},  \citet{1979lccs.book.....L, 1979nlcs.book.....L}, \citet{1991STIA...9233932G}, and the concluding spectroscopic studies of \cite{1991ApJS...77..417K} and the RECONS Research Consortium of Nearby Stars (\citealt{1995AJ....110.1838R}; \citealt{1997AJ....113.1458H}) at the end of the millennium, M dwarfs were relatively forgotten in the first decade of the current century. This apparent falling into oblivion, apart from a few honorable exceptions  (e.g., \citealt{2003AJ....126..921L}; \citealt{2005AJ....129.1483L}), was mostly due to many stellar astronomers focusing on the search and characterization of cooler objects with later spectral types: L, T, and, quite recently, Y (e.g., \citealt{1997A&A...327L..29M,1999AJ....118.2466M}, \citealt{1999ApJ...519..802K}; \citealt{2002ApJ...564..421B}; \citealt{2011ApJ...743...50C}).

Almost twenty years later, there have been a rebirth of M-dwarf studies. 
A few examples in the last four years are the possible existence of exoplanets in habitable zones around M dwarfs (\citealt{2009Natur.462..891C}; \citealt{2012ApJ...751L..16A}; \citealt{2013A&A...549A.109B}), a high occurrence of Earth-like-radius transiting exoplanets around cool stars in the {\em Kepler} field (\citealt{2010ApJ...713L.109B}; \citealt{2012ApJS..201...15H}; \citealt{2012ApJ...750L..37M}) and close to the Sun (\citealt{2010PASP..122..156A}; \citealt{2010PASP..122..149J}; \citealt{2011A&A...528A.111B}), transmission spectra of super-earths around M dwarfs (\citealt{2010ApJ...716L..74M}; \citealt{2012AJ....144..145B}), the luminosity and mass functions of low-mass stars in the solar neighborhood from Sloan data (\citealt{2010AJ....139.2679B}; \citealt{2011AJ....141...97W}), low contrast ratios that favor the detection of very faint, close-in (planetary) companions (\citealt{2010A&A...509A..52C}), high-precision dynamical masses of very low-mass binaries (\citealt{2010ApJ...711.1087K}), the complete new field of M-dwarf metallicity (\citealt{2010A&A...519A.105S}; \citealt{2010ApJ...720L.113R, 2012ApJ...748...93R}), fragile low-mass binaries (\citealt{2010MNRAS.404.1952B}; \citealt{2010AJ....139..176F}; \citealt{2010AJ....139.2566D}), or even a rebith of activity studies in light of new magnetohydrodynamic models (\citealt{2010MNRAS.407.2269M}; \citealt{2010AJ....139..504B}). 
Many exoplanet hunters turn now their eyes to M dwarfs, both with current instruments (CRIRES: \citealt{2010Msngr.140...41B}; NIRSPEC: \citealt{2010ApJ...723..684B}; MEarth: \citealt{2011ApJ...742..123I}, \citealt{2013ApJ...775...91B}; APOGEE: \citealt{2013AJ....146...81Z}; {\em Kepler}: \citealt{2013A&A...555A.108M}) and with future ones (SPIRoU: \citealt{2011ASPC..448..771A}; CARMENES: \citealt{2012SPIE.8446E..0RQ}; HPF: \citealt{2012SPIE.8446E..1SM}; {\em TESS}: \citealt{2010AAS...21545006R}; {\em EChO}: \citealt{2012ExA....34..311T}). 
In parallel, many research teams now focus on searching for the best M dwarfs for radial-velocity and transit exoplanet surveys (e.g., \citealt{2010ApJ...710..432R}; \citealt{2011AJ....142..138L}; \citealt{2013AJ....145..102L}; \citealt{2013MNRAS.435.2161F}; \citealt{2013prpl.conf2K020C}), apart from characterizing in detail such potential targets.

\begin{figure*}
	\centering
	\includegraphics[width=0.49\textwidth]{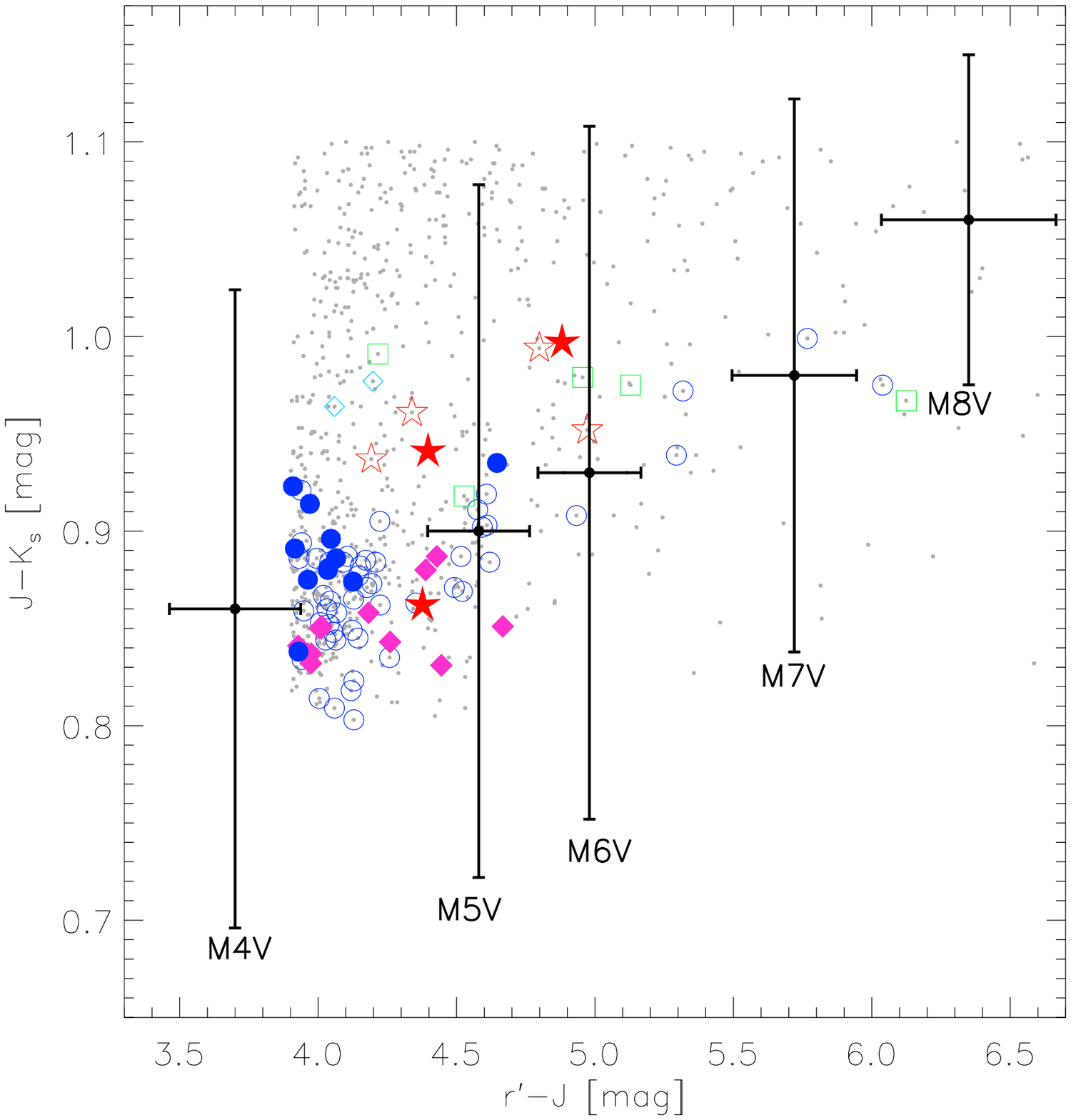}
	\includegraphics[width=0.49\textwidth]{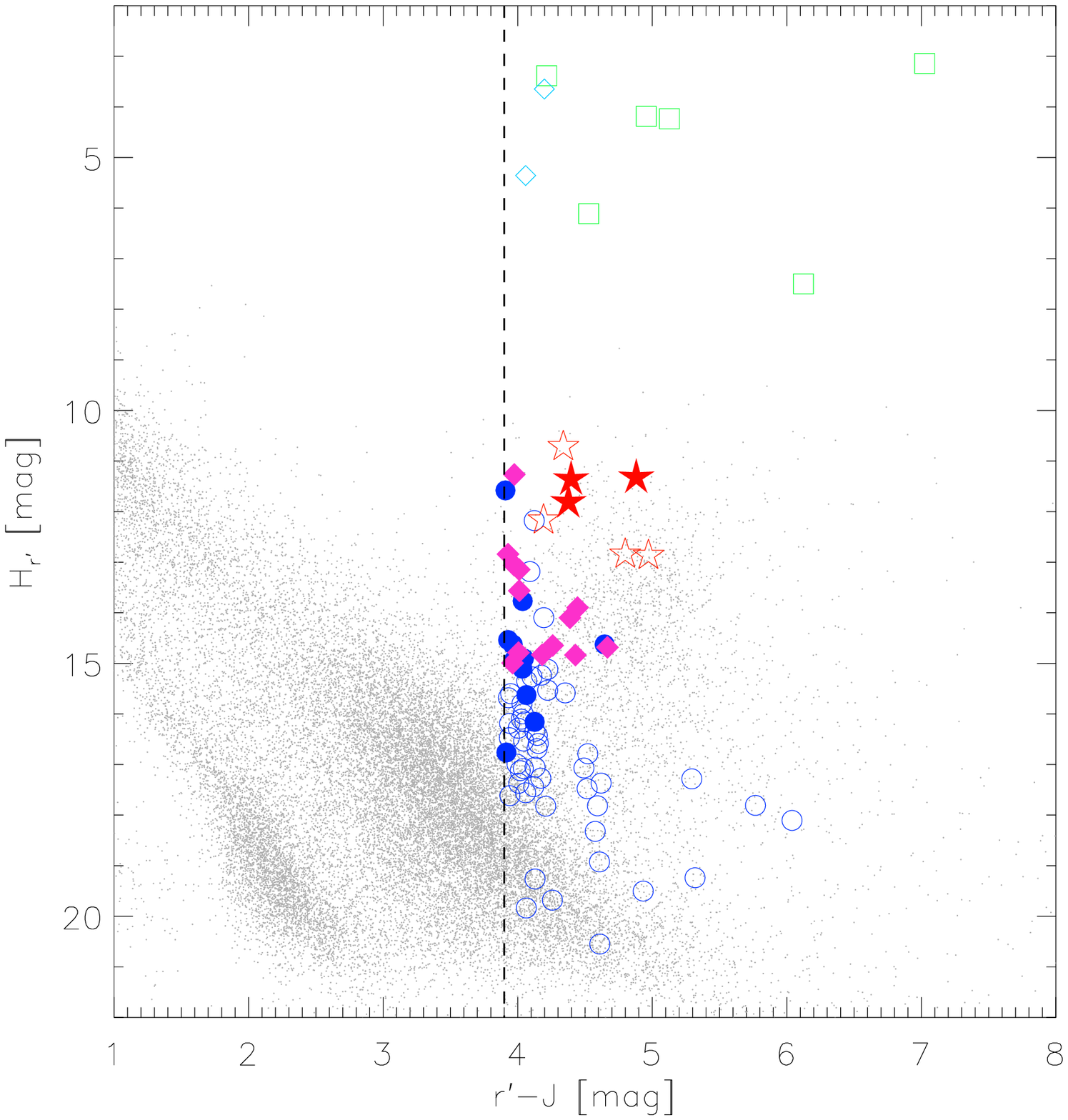}
	\caption{{\em Left panel:}
$J-K_{s}$ vs. $r'-J$ color-color diagram. 
Small gray solid points represent the full cross-matched sources. 
Filled symbols are for stars spectroscopically investigated in this work, while open symbols are for the cross-matched known stars in Table~\ref{table.known}.
New field M dwarfs reported here for the first time are drawn with (magenta) rhombs, previously known field M dwarfs with (blue) circles, members in Taurus-Auriga with (red) stars, M giants with (green) squares, and reddened Cygnus~OB2 massive stars with (cyan) rhombs.
Black error bars represent the average colors of typical dwarfs of spectral types M4\,V to M8\,V (\citealt{2008AJ....135..785W}).
{\em Right panel:}
same as left panel but the tiny gray points represent stars in \cite{2005AJ....129.1483L} with the CMC14 counterpart.
The vertical dashed line at $r'-J$ = 3.9\,mag indicates our color cut.
The two Galactic disk dwarf and halo subdwarf sequences are distinguishable in the bottom left corner.} 
\label{figure.diagrams}
\end{figure*}

In this work we search for unidentified bright intermediate M dwarfs in the solar neighborhood using Virtual Observatory (VO\footnote{\tt http://www.ivoa.net}) techniques. 
The VO is {\em ``an international initiative designed to provide the astronomical community with the data access and the research tools necessary to enable the exploration of the digital, multi-wavelength universe that is resident in the astronomical data archives''}. 
The VO is already an operational research infrastructure as demonstrated by the growing number of papers using VO tools (see, for instance, \citealt{2009mavo.proc....3C}; \citealt{2009A&A...497..973V}; \citealt{2011A&A...534L...7A}; \citealt{2012A&A...542A.105L}; \citealt{2013ApJ...767L...1L}; \citealt{2013A&A...556A.144L}; \citealt{2013ApJ...762...88M}, or \citealt{2013MNRAS.431.2063S} for recent examples of VO science papers focused on low-mass stars).
Besides, in this work we combine our VO search with a low-resolution spectroscopic follow-up, an astrometric and photometric study, and an activity analysis (based on our H$\alpha$ measurements and X-ray emission from public databases) to successfully identify not only potential targets for exoplanet hunting at less than 20\,pc, but also to serendipitously identify three young very low-mass stars  in the Taurus-Auriga region.

\section{Observations and Analysis}
\label{section.analysis}

\subsection{Candidate Selection}
\label{section.searchingcriteria}

%\begin{center}
\begin{scriptsize}
\begin{longtable*}{lcccc}   
      \caption{The 97 cross-matched objects that passed the final filters and were not spectroscopically analyzed}
%      \label{table.known}
\\ 
            \hline
  
Name                    & $\alpha$ (J2000)      & $\delta$ (J2000)      & SpT           & Class                 \\
                               & hh mm ss                    &  dd mm ss                                     &                       &                       \\                           
         
            \hline
            \endfirsthead
\multicolumn{5}{c}{{\bf \tablename\ \thetable{}.} Continued.}\\
\hline
  
Name                    & $\alpha$ (J2000)      & $\delta$ (J2000)      & SpT           & Class                 \\
                                        &                                       &                                       &                       &                       \\                           
             \hline
 \endhead   
 \endfoot           
 \\                                                                              
\endlastfoot      
LP 404--33              & 00 08 53.92                   & +20 50 25.2           & M4.5          & Dwarf         \\ % Alpha
TYC 2268-648-1       & 00 13 22.97                     & +33:47:02.2             & ...                     & Giant? \\
GJ 1011                 & 00 23 28.03                   & +24 18 24.4           & M4.0          & Dwarf         \\ % Gamma
G 130--68                       & 00 24 34.78                   & +30 02 29.5           & M4.5          & Dwarf         \\ % Beta
G 132--25 AB            & 00 45 56.63                   & +33 47 11.0           & M4.5+ & Dwarf         \\ % ...
G 69--32                        & 00 54 48.03                   & +27 31 03.6           & M4.5          & Dwarf         \\ % ...
IX And                  & 01 01 40.56                   & +37 53 46.5           & M4.0:         & Giant         \\ % ---
LSR J0155+3758  & 01 55 02.30                   & +37 58 02.8           & M5.0          & Dwarf         \\ % ...
2MASS J02132062+3648506         & 02 13 20.63                   & +36 48 50.7           & M4.5          & Dwarf         \\ % Alpha
LP 245--10              & 02 17 09.93                   & +35 26 33.0           & M5.0          & Dwarf         \\ % Alpha
FBS L 14-14             & 02 35 41.64                   & +26 03 03.2           & M6.5:         & Giant         \\ % ---
G 36--26                        & 02 36 44.13                   & +22 40 26.5           & M5.0          & Dwarf         \\ % ...
TYC 1779-1379-1    & 02 36 31.24                      & +29 35 55.7            & ....                     & Giant?\\
LSPM J0256+2359 & 02 56 13.96                   & +23 59 10.5           & M5.5          & Dwarf         \\ % Alpha
LP 355--27              & 03 07 46.82                   & +24 57 55.6           & M4.5          & Dwarf         \\ % ...
G 6--7                  & 03 26 44.96                   & +19 14 40.3           & M4.5          & Dwarf         \\ % Gamma
RX J0332.6+2843 & 03 32 35.79                   & +28 43 55.5           & M4.0          & Dwarf         \\ % Alpha
XEST 16--045            & 04 20 39.18                   & +27 17 31.7           & M4.0:         & Young         \\ % ...
GJ 1070                 & 04 22 33.49                   & +39 00 43.7           & M5.0          & Dwarf         \\ % Beta
G 8--31                 & 04 22 59.26                   & +25 59 14.8           & M4.0          & Dwarf         \\ % Beta
FW Tau AB               & 04 29 29.71                   & +26 16 53.2           & M5.5e+                & Young         \\ % ...
V546 Per                        & 04 30 25.27                   & +39 51 00.0           & M4.5          & Dwarf         \\ % Alpha
V927 Tau AB             & 04 31 23.82                   & +24 10 52.9           & M4.5e+                & Young         \\ % ...
G 8--41                 & 04 33 33.93                   & +20 44 46.2           & M4.0          & Dwarf         \\ % Beta
LP 415-1644              & 04 37 21.91                     &+19 21 17.4             & ...                      &  Dwarf? \\
Haro 6--36              & 04 43 20.23                   & +29 40 06.0           & M5.0e         & Young         \\ % ...
RX J0447.2+2038 & 04 47 12.25                   & +20 38 10.9           & M4.5          & Dwarf         \\ % ...
GJ 1072                 & 04 50 50.83                   & +22 07 22.5           & M5.0          & Dwarf         \\ % Alpha
LSPM J0501+2237 & 05 01 18.03                   & +22 37 01.6           & M4.5          & Dwarf         \\ % ...
HD 285190 BC            & 05 03 05.63                   & +21 22 36.2           & M4.5+ & Dwarf         \\ % Gamma
IRAS 05090+2027   & 05 12 03.98                     &+20 30  53.7             & ....                    &Giant? \\
{[LH98]} 190                  & 05 24 25.72                     &+19 22 07.0              & ...                      & Dwarf?\\
V780 Tau A              & 05 40 25.71                   & +24 48 09.0           & M5.5          & Dwarf         \\ % Gamma
LHS 6097                        & 05 58 53.33                   & +21 21 01.1           & M4.5          & Dwarf         \\ % ...
G 98--52 A              & 06 11 55.99                   & +33 25 50.6           & M3.5          & Dwarf         \\ % ...
0628+2052                 & 06 28 35.97                    &+20 52 37.6              & ....                     & Giant?\\
IRAS 06386+2330    & 06 41 41.14                    &+23 27 41.4                &  ...                       & Giant?\\ 
IRAS 06562+2229     &06 59 18.80                     &+22 24 53.4              & ...                      & Giant?\\
GJ 1093                 & 06 59 28.69                   & +19 20 57.7           & M5.0          & Dwarf         \\ % Alpha
GJ 1096                 & 07 16 18.02                   & +33 09 10.4           & M4.0          & Dwarf         \\ % Beta
BD+20 2091              & 08 28 24.90                   & +19 35 44.5           & M6.0:         & Giant         \\ % ---
DX Cnc                  & 08 29 49.34                   & +26 46 33.7           & M6.0          & Dwarf         \\ % Alpha
CV Cnc BC               & 08 31 37.44                   & +19 23 49.5           & M4.0+ & Dwarf         \\ % Gamma
$\rho$ Cnc B            & 08 52 40.85                   & +28 18 58.9           & M4.0          & Dwarf         \\ % Alpha
EI Cnc AB                       & 08 58 15.19                   & +19 45 47.1           & M5.5+ & Dwarf         \\ % Gamma
LP 368--128             & 09 00 23.59                   & +21 50 05.4           & M6.5          & Dwarf         \\ % Alpha
2MASS J09165078+2448559         & 09 16 50.78                   & +24 48 56.0           & M4.5          & Dwarf         \\ % ...
DX Leo B                        & 09 32 48.27                   & +26 59 44.3           & M5.5          & Dwarf         \\ % Beta
Ross 92                 & 09 41 02.00                   & +22 01 29.2           & M4.5          & Dwarf         \\ % Beta
LHS 2206                        & 09 53 55.23                   & +20 56 46.0           & M4.5          & Dwarf         \\ % Alpha
BD+25 2207              & 10 09 27.78                   & +24 57 31.8           & M5.0:         & Giant         \\ % ---
TYC 1968-490-1       &  10 13 52.77                    & +22 56 48.5             & ....                    & Giant?\\
G 54--19                        & 10 14 53.16                   & +21 23 46.4           & M4.5          & Dwarf         \\ % ...
LP 374--39              & 11 23 08.00                   & +25 53 37.0           & M5.0          & Dwarf         \\ % Beta
G 121--28                       & 11 52 57.91                   & +24 28 45.4           & M4.5          & Dwarf         \\ % Beta
G 148-47 B                 & 12 21 26.69                     & +30 38 37.0             & ...                     & Dwarf?\\ 
Sand 58 A               & 12 21 27.05                   & +30 38 35.7           & M5.0          & Dwarf         \\ % ...
HD 108421 C             & 12 26 57.37                   & +27 00 53.7           & M4.5          & Dwarf         \\ % ...
G 123--45                       & 12 36 28.70                   & +35 12 00.8           & M3.5          & Dwarf         \\ % ...
LP 377--36              & 12 39 43.54                   & +25 30 45.7           & M4.5          & Dwarf         \\ % ...
Sand 214                        & 13 06 50.25                   & +30 50 54.9           & M5.0          & Dwarf         \\ % Beta
GJ 1167 A               & 13 09 34.95                   & +28 59 06.6           & M4.0          & Dwarf         \\ % Alpha
EK CVn                  & 13 14 32.49                   & +34 20 55.9           & M6.0:         & Giant         \\ % ---
GJ 1171                 & 13 30 31.06                   & +19 09 34.0           & M4.5          & Dwarf         \\ % ...
LP 323--169             & 13 32 39.09                   & +30 59 06.6           & M4.5          & Dwarf         \\ % ...
GJ 1179 A               & 13 48 13.41                   & +23 36 48.6           & M5.5          & Dwarf         \\ % Alpha
G 166--33                       & 14 29 59.56                   & +29 34 02.9           & M4.0          & Dwarf         \\ % Gamma
NLTT 39916               &  15 19 21.23                    & +34 03 42.8             & ...                      & Dwarf?\\
G 167--47                       & 15 31 54.27                   & +28 51 09.6           & M4.5          & Dwarf         \\ % Beta
G 180--11 AB            & 15 55 31.78                   & +35 12 02.9           & M4.5+ & Dwarf         \\ % Gamma
HD 190360 B             & 20 03 26.52                   & +29 52 00.0           & M4.5          & Dwarf         \\ % Beta
J2006+3651             &  20 06 54.00                    &+36 51 48.3              & ...                     & Giant?\\
J2011+3423              & 20 11 56.42                     &+34 23 58.1               & ...                     &Giant?\\
J2014+3943              & 20 14 55.51                     & +39 43 26.7             & ...                          & Dwarf?\\ 
HD 346301A              & 20 19 02.83                     & +22 05 21.4             & ...                    & Giant?\\
J2022+4030              & 20 22 14.67                     & +40 30 02.7              & ...                        &  Giant?\\  
J2023+2451              & 20 23 34.27                     &+24 51 20.1              & ...&  Giant?\\  
J2024+3930              & 20 24 38.71                    &+39 30 30.1               & ...&  Giant?\\  
J2026+3733              & 20 26 16.44                     &+37 33 01.2              & ...&  Giant?\\  
G 210-20                    & 20 28 22.08                     & +34 12 08.7              & ... & Dwarf?\\
{[CPR2002] A25} & 20 32 38.44                   & +40 40 44.5           & O8\,III               & Young \\ % ---
J2032+4042               & 20 32 43.88                     &+40 42 17.1              & ... & Giant?\\
{[CPR2002] A20} & 20 33 02.92                   & +40 47 25.4           & O8\,II((f))           & Young \\ % ---
{[CPR2002] A22}          & 20 33 11.29                    & +40 42 33.7            &  ...    & Giant?\\
G 210-26                     &  20 33 15.77                    &+28 23 44.0              & ...   &Dwarf?\\
J2038+4021               & 20 38 17.13                    &+40 21 06.0                & ... & Giant?\\
LSPM J2045+3508  & 20 45 22.16                     &+35 08 15.1                & ...&  Dwarf?\\ 
G 211--9                        & 21 02 46.06                   & +34 54 36.0           & M4.5          & Dwarf         \\ % ...
IRAS 21044+2818  & 21 06 35.76                      &+28 31 06.1            & ...  &  Giant?\\
V445 Vul                        & 21 08 01.33                   & +23 43 44.6           & M7.0:         & Giant         \\ % ---
TYC 2710-1557-1    &  21 09 13.22                     &+34 08 48.0             & ...                     & Giant?\\
LSR J2124+4003  & 21 24 32.34                   & +40 04 00.0           & M6.5          & Dwarf         \\ % Alpha
BD+29 4448               &   21 33 57.94                   &+29 49 11.4             & ...                     & Giant?\\
J2154+1914                &  21 54 51.21                   &+19 14 10.7               & ...                     & Giant?\\
G 130-31                       & 23 59 19.80                    &+32 41 23.7             & ...                      & Dwarf?\\
G 127--50                       & 22 43 23.13                   & +22 08 17.9           & M4.5          & Dwarf         \\ % Gamma
GJ 1288                 & 23 42 52.74                   & +30 49 21.9           & M4.5          & Dwarf         \\ % Alpha
\hline
\label{table.known}
\end{longtable*}
\begin{minipage}{14cm}
{$^{a}$}{Dwarf: known dwarf; Giant: known giants; Young: known young stars (T Tauri and reddened massive stars); Dwarf?: probable dwarf; Giant?: probable giants}
\end{minipage}
\end{scriptsize}
%\end{center}

%__________________________________________________ Two column table
  \begin{table*}
     \scriptsize
\renewcommand\tabcolsep{1.8pt}
      \caption[]{Basic Data of the 27 Spectroscopically Analysed M Candidates$^{a}$}       
         \label{table.basicdata}
     $$ 
         \begin{tabular}{ll cc ccc cc c}
            \hline
            \hline
            \noalign{\smallskip}
            ID			& Alternative         	& $\alpha$ (J2000)		& $\delta$ (J2000) 		& $b$	             	& $\mu_{\alpha}\cos{\delta}$	& $\mu_{\delta}$		& $r'-J$ 		& $J-K_s$		&$H_{r'}$\\
     	                            & Name               	&                   	                   &     	     		               	& (deg)                      	& (mas\,a$^{-1}$)              	          	& (mas\,a$^{-1}$)     		& (mag)		& (mag)		& (mag)	\\ 
            \noalign{\smallskip}
            \hline
            \noalign{\smallskip}
 J0012+3028	       & ...	         	                             		& 00:12:13.43 		& +30:28:44.2		&$-$31.6 		& +52.7$\pm$5.1   		& $-$25.9$\pm$5.1  	    	&3.97 &         0.83				&  13.06		\\
 J0013+2733	       & ...	         				 		& 00:13:19.52 		& +27:33:30.8 		&$-$34.6		& +16.3$\pm$4.4   		& $-$116.4$\pm$4.4	     	&4.00 &       0.85  				&  14.79		\\
 J0024+2626	       & \object{LSPM J0024+2626}           	& 00:24:03.81 		& +26:26:29.7 		&$-$36.0      	& +7.0$\pm$6.4    		& $-$30.1$\pm$6.4       	&3.90 &         0.92 				&  11.58		\\
 J0058+3919 	       & \object{PM I00580+3919} 	          & 00:58:01.13 		& +39:19:11.2 		&$-$23.5		& $-$102.6$\pm$9.3         & +34.1$\pm$9.3  		&4.04 &         0.88				&   13.77		\\
 J0122+2209	       & \object{G 34$-$23}	 	          	& 01:22:10.32 		& +22:09:03.0 	        	&$-$40.2		& +237.0$\pm$5.3  		& $-$152.7$\pm$5.3		&3.96  &         0.87  				&  14.63  	\\
 J0156+3033           & \object{NLTT 6496} (K\"o4A)	           & 01:56:45.76 		& +30:33:28.8 		&$-$30.2		& +219.2$\pm$4.6  		& $-$12.5$\pm$4.6  		&4.13  &         0.87 				&  16.16  	\\
 J0304+2203	       & ...	         	                             		& 03:04:44.10 		& +22:03:21.2 		&$-$31.2		& +37.8$\pm$5.1   		& $-$49.3$\pm$5.1		&4.44 &         0.83				&  13.90		\\
 J0326+3929	       & \object{LSPM J0326+3929W}       	& 03:26:34.20 		& +39:29:02.5 		&$-$14.2		& +61.4$\pm$5.7    		& --144.1$\pm$5.7   		& 4.00 &         0.91				&   14.94	\\
 J0327+2212	       & ...	         	                             		& 03:27:30.87 		& +22:12:38.1 		&$-$27.9 		& $-$39.0$\pm$6.3   	& $-$53.0$\pm$6.3 	         	&4.01 &         0.85			&  13.15		\\
 J0341+1824	       & ...	         	                             		& 03:41:43.87 		& +18:24:06.2 		&$-$28.6		& +23.9$\pm$4.7   		& $-$42.2$\pm$4.7 		&3.93   &      0.84   				&  12.84	\\
 J0342+2326	       & \object{LR Tau	}                    		& 03:42:53.29 		& +23:26:49.5 		&$-$24.6		& +176.7$\pm$4.7  		& $-$61.2$\pm$4.7  		&4.06  &         0.89 				& 15.63		\\
 J0422+2439	       & ...	         		          			& 04:22:54.17 		& +24:39:53.6 		&$-$17.3		& +5.0$\pm$4.5	      	& $-$22.3$\pm$4.5  		&4.88    &    1.00 				&   11.32		\\
 J0424+3706	       & ...	         						& 04:24:21.51 		& +37:06:20.8 		&$-$8.6		& +47.3$\pm$6.5   		& $-$57.6$\pm$6.5 		&4.01     &    0.85   				&  13.56		\\
 J0435+2523	       & ...	         						& 04:35:47.79 		& +25:23:43.6 		&$-$14.6		& +2.9$\pm$4.5         		& $-$21.6$\pm$4.5  		&4.40 &         0.94				&  11.36		\\
 J0439+2333	       & ...	         		         				& 04:39:04.54 		& +23:33:19.9 		&$-$15.3		& $-$0.9$\pm$4.6     		& $-$24.6$\pm$4.6  		&4.38 &         0.86 				&  11.81		\\
 J0507+3730	       & ...	         						& 05:07:14.45 		& +37:30:42.1 		&$-$1.9		& $-$105.5$\pm$5.0  	& $-$8.0$\pm$5.0 		&4.43 &         0.89				&  14.84		\\
 J0515+2336	       & ...	         						& 05:15:17.54 		& +23:36:25.9 		&$-$8.6		& +36.7$\pm$4.6   		& $-$71.5$\pm$4.6  		&4.39 &         0.88 				&  14.10	\\
 J0630+3003	       & ...	         						& 06:30:10.18 		& +30:03:39.5 		&+9.0		& $-$1.6$\pm$8.2    		& +28.1$\pm$8.2   		&3.97 &        0.84   				&  11.27		\\
 J0909+2247	       & ...	         						& 09:09:07.97 		& +22:47:41.2 		&+39.8		& $-$81.2$\pm$4.2   	& $-$71.8$\pm$4.2 		&4.18 &         0.86  				&  14.83		\\
 J1132+1816	       & ...	         						& 11:32:23.00 		& +18:16:22.4 		&+69.8		& +136.4$\pm$5.5  		& +58.6$\pm$5.5  		&3.96 &         0.84 				&  14.99		\\
 J1241+1905	       & \object{G 59$-$34}	 	         		& 12:41:29.00 		& +19:05:00.7 		&+81.6		& +68.6$\pm$4.9   		& $-$305.4$\pm$4.9		&3.92 &         0.89				&  16.76	\\ 
 J1459+3618	       & \object{RX~J1459.4+3618}       	& 14:59:25.04 		& +36:18:32.3 		&+61.4		& $-$123.9$\pm$5.2 		& +76.1$\pm$5.2       	&4.03 &         0.88				&  15.10		\\
 J1518+2036	       & ...	         						& 15:18:31.45 		& +20:36:28.3 		&+55.9		& +10.1$\pm$4.6   		& +94.8$\pm$4.6  		&4.67 &         0.85				&  14.68 		\\
 J1547+2241	       & \object{LSPM J1547+2241}         	& 15:47:40.69 		& +22:41:16.5 		&+50.0		& $-$180.9$\pm$5.0  	& $-$29.5$\pm$5.0 		&4.05  &        0.90   				&  14.90		\\
 J2211+4059	       & \object{1RXS J221124.3+410000} 	& 22:11:24.14 		& +40:59:58.9 		&$-$12.4		& $-$89.6$\pm$4.9   	& +68.0$\pm$4.9 		&4.65 &         0.93 				&  14.63		\\
 J2248+1819	       & \object{PM I22489+1819} 		& 22:48:54.58 		& +18:19:58.9 		&$-$35.7		& $-$24.7$\pm$5.1   	& $-$132.8$\pm$5.1		&3.93 &         0.84				&   14.54		\\
J2259+3736	       & ...	         						& 22:59:14.81 		& +37:36:39.6 		&$-$20.1		& +96.2$\pm$4.7   		& $-$28.9$\pm$4.7 		&4.26  &        0.84   				&  14.65		\\
         \noalign{\smallskip}
            \hline
         \end{tabular}
     $$ 
\begin{minipage}{18cm}
{$^{a}$}{Identification, discovery name (blank if new), right ascension and declination from 2MASS, Galactic latitude, proper motions from PPMXL, $r'-J$ and $J-K_s$ colors, and reduced proper motion $H_{r'} = r' + 5 \log{\mu} + 5$.}
\end{minipage}
   \end{table*}

First of all, we cross-matched the whole  Carlsberg Meridian Catalogue 14 (CMC14\footnote{\tt http://www.ast.cam.ac.uk/$\sim$dwe/SRF/cmc14.html}; \citealt{2006yCat.1304....0C} -- see \citealt{2002A&A...395..347E} for a description of a previous release) with the Two-Micron All Sky Survey (2MASS; \citealt{2006AJ....131.1163S}).
This correlation was performed with the help of the Aladin sky atlas (\citealt{2000A&AS..143...33B}) and the Starlink Tables Infrastructure Library Tool Set (STILTS; \citealt{2006ASPC..351..666T}).
To avoid memory overflow problems, we divided the {25\,078}\,deg$^2$ of common CMC14-2MASS sky (i.e., the CMC14 area -- over 60\,\% of the whole sphere) into overlapping circular regions of 30\,arcmin radius.
We used a matching radius of 5\,arcsec, which ensured that objects with high proper motions of up to $\mu \sim$ 1700\,mas\,a$^{-1}$ (given the typical baseline between 2MASS and CMC14 astrometric epochs) were not left out.
Only the closest 2MASS counterpart to each CMC14 source was considered. 

Constraints based on the colors expected for M dwarfs with spectral types M4\,V and later were imposed. In particular, we selected objects with colors $r'-J >$ 3.9\,mag and 0.8 mag $< J-K_{\rm s} <$ 1.1\,mag (e.g., \citealt{2008AJ....135..785W}) and high quality 2MASS flags (AAA). 

Since we were interested in {\em bright} M dwarfs, our last restriction was $J <$ 10.5\,mag.
We did not expect to identify any dwarfs later than M8\,V ($r'-J \gtrsim$ 6.5\,mag) because the CMC14 completeness magnitude is $r' \approx$ 17.0\,mag in the Sloan passband (slightly variable by 0.1--0.2\,mag from one sky region to another).
After these color and magnitude cuts, we expect  to find M4\,V (M8\,V) stars within 45\,pc  (7.5\,pc)  the Sun using the $M_J$-SpT relation in Caballero et al. (2008,  Table 3).
The selection of the resulting 828 sources with $r'JHK_{\rm s}$ photometry is illustrated in the left panel of Figure.~\ref{figure.diagrams}.

We prepared a list of high-priority targets for a spectroscopic run planned for the Canarian winter (see Section~\ref{section.spectroscopy}).
Of the 828 stars, we selected objects visible during the season (i.e., with right ascensions 22\,h $< \alpha <$ 16\,h) and with minimum zenith distances during culmination (i.e., with declinations +18\,deg $< \delta <$ +41\,deg).
We imposed an extra cut at colors $J-K_{\rm s} <$ 1.0\,mag, thus making the near-infrared color constraint actually 0.8 mag$< J-K_{\rm s} <$ 1.0 mag.
The 125 sources passing these filters  were inspected visually with Aladin and classified into five groups: 
(1) known dwarfs, (2) known giants, (3) known young stars (T~Tauri and reddened massive stars), 4) probable giants, and (5) probable dwarfs. 
There was only one artifact from an incorrect CMC14-2MASS cross-match of a visual binary, which resulted in a final list of 124 stars.
For the classification, we used additional information gathered from the SIMBAD\footnote{\tt http://simbad.u-strasbg.fr/simbad/sim-fid} and ADS\footnote{\tt http://adswww.harvard.edu/} services, and the PPMXL catalog of positions and proper motions on the ICRS (PPMXL; \citealt{2010AJ....139.2440R}) and {\em IRAS} point source (\citealt{1988iras....7.....H}).
Field giants in the investigated magnitude range have in general very low proper motions, lower than 5\,mas\,a$^{-1}$, near-infrared colors close to the upper limit ($J-K_{\rm s} \approx$ 1.0\,mag), flux excess in the {\em IRAS} passbands, and, in some cases, photometric variability due to pulsations (see, e.g., the recent VO-based survey for bright Tycho-2 stars with red colors by \citealt{2012A&A...539A..86J}). 

The 97 cross-matched known dwarfs, giants and young stars with spectral-type determination and probable dwarfs and giants without spectral typing are shown in Table~\ref{table.known}.
All the known dwarfs (55) except one have spectral types M4.0\,V or later (the exception is \object{G~98--52}~A, which has an M3.5\,V spectral type).
Table~\ref{table.known} includes four pre-main sequence T~Tauri stars in Taurus-Auriga (\object{XEST~16--045}, \object{FW~Tau}~AB, \object{V927~Tau}~AB, and \object{Haro~6--36}; see Section~\ref{section.taurus}) and a couple of reddened massive stars in Cygnus~OB2 (\object{[CPR2002]~A20} and \object{[CPR2002]~A25}; \citealt{2002A&A...389..874C}).
Field M giants, T~Tauri stars, and reddened Cygnus~OB2 massive stars fall in well-defined locations in a reduced-proper-motions diagram, as the one shown in the right panel of Figure.~\ref{figure.diagrams}.

The remaining 27 high-priority dwarf candidates, shown in Table~\ref{table.basicdata}, were selected for spectroscopic follow-up.
They all had proper motions greater than 20\,mas\,a$^{-1}$, no {\em IRAS} detection, and magnitudes and colors consistent with intermediate or late M spectral type and luminosity class V. 
Of them, only 11 were previously identified and classified as M dwarf
candidates based only on photometry by \cite{1959LowOB...4..136G, 1961LowOB...5...61G}, \cite{1964RA......6..535M}, \cite{1979nlcs.book.....L}, \cite{1998ApJ...504..461F}, \cite{2005AJ....129.1483L}, and \cite{2011AJ....142..138L}.

\subsection{Spectroscopy}
\label{section.spectroscopy}

%______________________________________________ Figure 
\begin{figure*}
	\centering
	\includegraphics[width=0.45\textwidth]{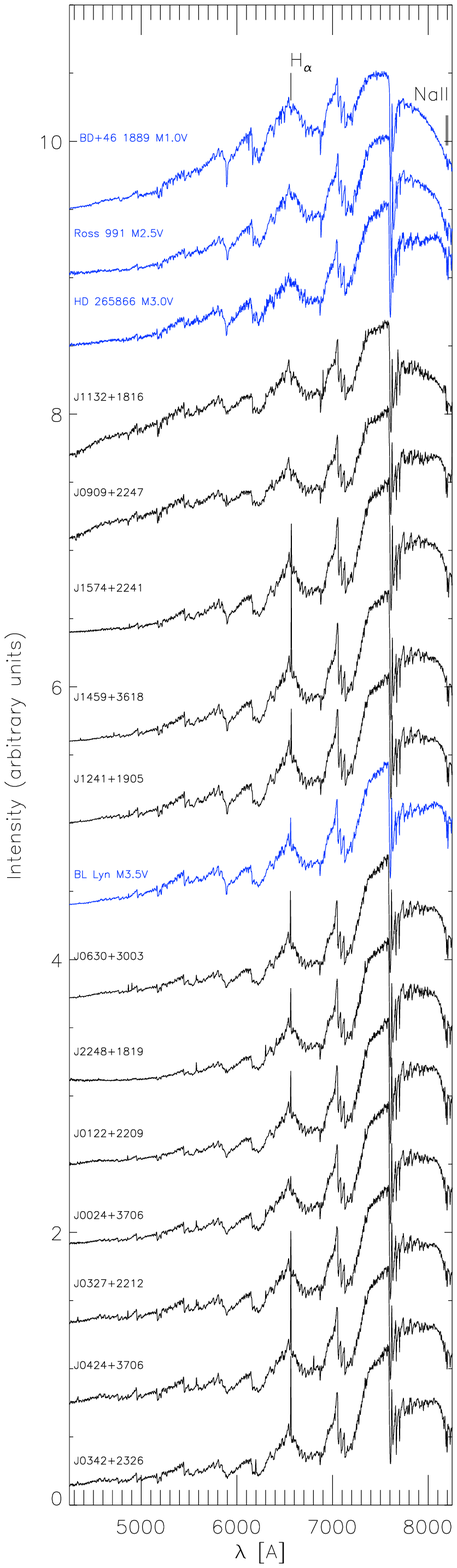}
	\includegraphics[width=0.45\textwidth]{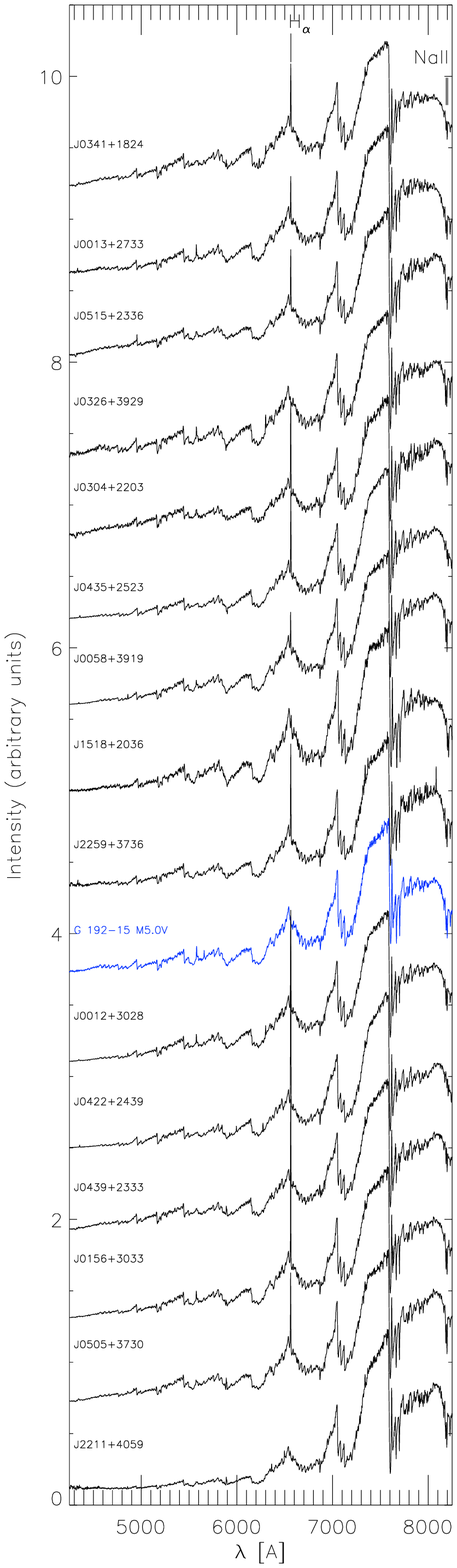}
	\caption{IDS/INT spectra of our 27 M dwarf candidates (in black) and five reference stars (in blue), normalized at 7400\,{\AA}.}
\label{figure.spectra}
\end{figure*}

On 2012 January 11--13, we used the Intermediate Dispersion Spectrograph (IDS) at the 2.5\,m Isaac Newton Telescope (INT) in the 
Observatorio del Roque de Los Muchachos (La Palma, Spain).
We used the configuration with the Red+2 detector, the R300V grating centered on 550\,nm, and the 1.0\,arcsec wide slit, which provided a resolution $\mathcal R \approx$ 1600 over a wide wavelength interval from 360 to 900\,nm.
The actual useful wavelength interval was, however, a bit narrower, from 425 to 825\,nm.

%__________________________________________________ One column table
   \begin{table}
   \scriptsize
   \renewcommand\tabcolsep{2.5pt}
      \caption[]{Spectral-type reference stars} 
         \label{table.referencestars}
     $$ 
         \begin{tabular}{l l cc cc}
            \hline
            \hline
            \noalign{\smallskip}
Name & GJ			& $\alpha$ (J2000)    	& $\delta$ (J2000)	& SpT		& SpT 		\\
        	  &				&        				& 	        			& PMSU$^{a}$		& Simbad 		\\	   	             
            \noalign{\smallskip}
           \hline
            \noalign{\smallskip}
\object{G 192--15} & 3380	& 06:02:29.18			& +49:51:56.2		& M5.0\,V	    	& M5.0\,V	 	\\
\object{HD 265866} & 251	& 06:54:48.96			& +33:16:05.4		& M3.0\,V       	& M4.0\,V		\\
\object{BL Lyn} & 277 B	& 07:31:57.33			& +36:13:47.4		& M3.5\,V		& M4.5\,V 		\\
\object{Ross 991} & 3748	& 12:47:00.99			& +46:37:33.4		& M2.5\,V      	& M2.0\,V	 	\\
\object{BD+46 1889} & 521& 13:39:24.10			& +46:11:11.4		& M1.0\,V	    	& M2.0\,V	 	\\
         \noalign{\smallskip}
            \hline
         \end{tabular}
         $$
         \begin{list}{}{}
\item[$^{a}$]The Palomar/Michigan State University survey (\citealt{1995AJ....110.1838R}; \citealt{1996AJ....112.2799H}).
\end{list}
        \end{table}

We collected low-resolution spectra of the 27 high-priority dwarf candidates and 5 reference stars for spectral-type determination, which are shown in Table~\ref{table.referencestars}.
Exposure times ranged between 700 and 2400\,s, depending on target brightness.
Raw data were next reduced following standard procedures within the IRAF environment (bias and flat-field correction, 
cosmic-ray rejection, and optimal extraction).
Wavelength calibration was carried out with spectra of Cu-Ne arc lamps taken during the run.
Figure.~\ref{figure.spectra} shows the 32 final spectra of bright M dwarf candidates and reference stars sorted by spectral type.

\section{Results}
\label{section.results}

\subsection{Spectral Types}
\label{spectraltypes}

%__________________________________________________ Two column table
   \begin{table*}
    \scriptsize
    \renewcommand\tabcolsep{5.5pt}
      \caption[]{Miscellaneous Data of the 27 Investigated Stars$^{a}$}
         \label{table.additional}
     $$ 
         \begin{tabular}{lcccccccclccc}
            \hline
            \hline
            \noalign{\smallskip}
 ID			& SpT    		& $\Re$	& SpT   	& pEW(H$\alpha$)$^b$  	& pEW(Na~{\sc i})	& $T_{\rm eff}$	& $\log{g}$	& $d^c$	           		& 1RXS     			&  $\rho$	& $X$-$CR$		& $HR1$  \\
         			&  (biblio.)      	&	& (IDS)      & [$\AA$]				& [$\AA$]  			&[K]             	& 		         & [pc]      				&          	     			& [arsec]	& [s$^{-1}$]       &         	\\	   	             
            \noalign{\smallskip}
            \hline
            \noalign{\smallskip}
 J0012+3028	& ...			&2.87	& M5.0\,V   	&$-$10.3$\pm$0.2  		& +4.5$\pm$0.2    	& 	3100  	& 	5.5   & 14$\pm$4	 			& J001213.6+302906 	& 21.9	& 0.02  & $-$0.08 \\
 J0013+2733	& ...			& 2.56	& M4.5\,V   	& $-$4.4$\pm$0.2    		& +6.1$\pm$0.2    	&  	3100  	          & 	5.5   & 20$\pm$5	 			& ...					& ...		& ...		& ...            	\\
 J0024+2626	& ...			&2.43	& M4.0\,V		& $-$2.0$\pm$0.2		& +6.3$\pm$0.2  	&  	3100  	          & 	{\bf 5.5}& 18$\pm$5					& ...					& ...		& ...		& ...              \\
 J0058+3919	& m5:\,V		&2.76	& M4.5\,V		& $-$3.4$\pm$0.3		& +5.0$\pm$0.3  	&  	3100  	& 	{\bf 5.5}& 14$\pm$3					& J005802.4+391912 	&14.8 	& 0.02	& $-$0.25   \\     
 J0122+2209	& M4.5\,V   	&2.43	& M4.0\,V		& $-$4.8$\pm$0.2		& +6.7$\pm$0.3  	&  	3100  	          & 	5.5& 8$\pm$2	 				& J012210.9+220909 	&10.4  	& 0.19	& $-$0.22  \\
 J0156+3033	& m4.5:\,V		&3.00	& M5.0\,V		& $-$9.3$\pm$0.3		& +5.4$\pm$0.3  	&  	3100  	           & 	{\bf 6.0}& 15$\pm$4					& J015645.8+303332 	& 3.25 	& 0.04 	& $-$0.54\\   
 J0304+2203	& ...			&2.62		& M4.5\,V		& $-$8.5$\pm$0.3		& +3.7$\pm$0.3  	&  	3000  	& 	5.5& 21$\pm$5	 				& J0304441.3+220320	& 30.0      	& 0.02       & +0.10 \\
 J0326+3929	& k7:\,V	&2.61		& M4.5\,V		& {$>$ --1.0} 			& +5.7$\pm$0.3  	&      3100 	& 	 4.5& 17$\pm$4	 			& ...					& ...		& ...		& ...               \\
 J0327+2212	& ...			& 2.46		& M4.0\,V   	& $-$3.7$\pm$0.1    		& +4.4$\pm$0.2    	&  	3100 	& 	5.5& 17$\pm$4	 				& ...					& ...		& ...		& ...  \\ 
 J0341+1824	& ...			& 2.53		& M4.0\,V   	& $-$6.0$\pm$0.2    		& +3.2$\pm$0.3    	&  	3100 	& 	5.5& 21$\pm$5     				& ...					& ...		& ...		& ...  \\
 J0342+2326	& m:\,V		&2.52		& M4.0\,V   	& $-$7.7$\pm$0.2    		& +6.4$\pm$0.3    	&  	3100  	& 	6.0& 18$\pm$5	 				& ...					& ...		& ...		& ...   \\
 J0422+2439	& ...			& 2.93	         & M5.0e   	         &$-$20.9$\pm$0.5		& +2.1$\pm$0.5	&  	2900  	& 	3.0& $\sim$140	 			& J042254.9+243950  	& 10.4   	& 0.03	& +1.0  \\        
 J0424+3706	& ...			& 2.49		& M4.0\,V		& $-$8.8$\pm$0.2    		& +4.7$\pm$0.3    	&  	3100 	& 	6.0& 18$\pm$5     	 			& J042421.4+370609 	& 11.4 	& 0.03  	& $-$0.66 \\ 
 J0435+2523	& ...			& 2.65		& M4.5   	         & $-$9.0$\pm$0.2     		& +2.7$\pm$0.5   	&  	3000  	& 	3.5& $\sim$140				& ...					& ...		& ...		& ...             \\
 J0439+2333	& ...			& 3.00		& M5.0   	          & $-$8.2$\pm$0.3              & +2.2$\pm$0.5    	&  	3000  	& 	4.0& $\sim$140				& ...					& ...		& ...		& ...               \\
 J0507+3730 & ...			& 3.02		& M5.0\,V       	& $-$7.4$\pm$0.4    		& +5.6$\pm$0.3    	&  	3000  	& 	5.5& 15$\pm$4    	 			& J050714.8+373103 	& 21.8 	& 0.04 	&$-$0.34     \\
 J0515+2336 & ...			& 2.59		& M4.5\,V   	& $-$5.3$\pm$0.2    		& +5.5$\pm$0.3     	&  	 3000 	& 	5.5& 18$\pm$5    	 			& ...					& ...		& ...		&  ...         \\
 J0630+3003 & ...			& 2.30		& M4.0\,V      	& $-$4.4$\pm$0.2   		& +5.7$\pm$0.2     	&  	3100  	& 	5.5& 22$\pm$5	 				& ...					& ...		& ...		&...                \\
 J0909+2247 & ...			& 1.98		& M3.0\,V       & {$>$ --1.0} 			& +4.6$\pm$0.2    	&  	3000  	& 	5.0& 44$\pm$11    		& ...					& ...		& ...		&...                \\  
 J1132+1816 & ...			& 1.84		& M2.5\,V       & {$>$ --1.0} 			& +2.2$\pm$0.3     	&  	3200        & 	 5.5& 39$\pm$10    		& ...					& ...		& ...		&...              \\
 J1241+1905 & M4.5\,V	& 2.23		& M3.5\,V       & $-$3.7$\pm$0.2  		& +5.6$\pm$0.3     	&  	3100  	& 	 5.5& 33$\pm$8 	 	& ...					& ...		& ...		& ...               \\
 J1459+3618 & ...			& 2.19		& M3.5\,V       & $-$8.4$\pm$0.2   		& +5.7$\pm$0.3    	&  	3100  	& 	 6.0& 31$\pm$8       				& J145924.6+361826 	& 8.24 	& 0.02 	& +0.04      \\
 J1518+2036 & ...			& 2.78		& M4.5\,V       & $-$1.1$\pm$0.1 		& +7.4$\pm$0.3    	&  	2900  	& 	5.0& 14$\pm$3      		& ...					& ...		& ...		& ...            \\
 J1547+2241 & m5:\,V        & 2.16		& M3.5\,V       & $-$3.9$\pm$0.1    		& +6.0$\pm$0.3   	&  3100  	& 	6.0& 22$\pm$5     	 		& J154741.3+224108 	& 11.7 	& 0.03	&$-$0.56    \\
 J2211+4059 & m7:\,V		& 3.30       	          & M5.5\,V       & $-$5.4$\pm$0.2    		& +7.1$\pm$0.3  	 &  	2900 	&       5.0  &  9$\pm$2      				& J221124.3+410000 	& 2.04	& 0.06	& +0.11  \\ 
 J2248+1819 & m5:\,V		& 2.42		& M4.0\,V       & $-$4.2$\pm$0.2   		& +7.1$\pm$0.3     	 &  	3100 	&        5.5 & 16$\pm$4     			& ...					& ...		& ...		&...             \\
 J2259+3736 & ...			& 2.81 	          & M4.5\,V       & $-$9.8$\pm$0.2   		& +5.9$\pm$0.3     	&  	3000 	&	 4.5	& 12$\pm$3     			& ...					& ...		& ...		& ...             \\
         \noalign{\smallskip}
            \hline
         \end{tabular}
     $$ 
 \begin{minipage}{18cm} 
{$^{a}$}{Spectral types from the bibliography and our IDS/INT spectra  (photometric spectral types are listed with `m' and `k'), $\Re$ index, pseudo-equivalent widths of H$\alpha$ $\lambda$656.3\,nm and Na~{\sc i} $\lambda\lambda$818.3,819.5\,nm from our spectra, $T_{\rm eff}$ ($\pm$100\,K) and $\log{g}$ ($\pm$0.5) from our VOSA fits, derived heliocentric distance, and key data from the {\em ROSAT} All-Sky Bright and Faint Survey Catalogues (1RXS name, angular separation between the X-ray and 2MASS coordinates, count rate, and hardness ratio).}\\
{$^{b}$}{Two stars had previous pEW(H$\alpha$) determinations: 
J0122+2209 of --4.1$\pm$0.7\,{\AA} and
J1459+3618 of --6.9$\pm$0.6\,{\AA} (\citealt{2002AJ....124.2868M}).}\\
{$^{c}$}{Three stars had previous distance determinations: 
J0122+2209 at 10.5\,pc and J1459+3618 at 22.0\,pc (\citealt{1998ApJ...504..461F}),
J0156+3033 at 19$^{+6}_{-4}$\,pc  \citep{2012Obs...132....1C}.}\\
  \end{minipage} 
   \end{table*}

As explained above, 11 of the 27  high-priority M-dwarf candidates observed spectroscopically were previously known.
Of them, seven had spectral-type estimations from colors (from digitization of blue and red photographic plates --\citealt{1961LowOB...5...61G}; \citealt{2011AJ....142..138L}-- or from optical and near-infrared multi-band photometry --\citealt{1998ApJ...504..461F}),  and two from real spectra (J0122+2209 from \citealt{2013AJ....145..102L} and J1241+1905 from \citealt{2003AJ....126.3007R}).

\begin{figure}
	\centering
	\includegraphics[width=0.49\textwidth]{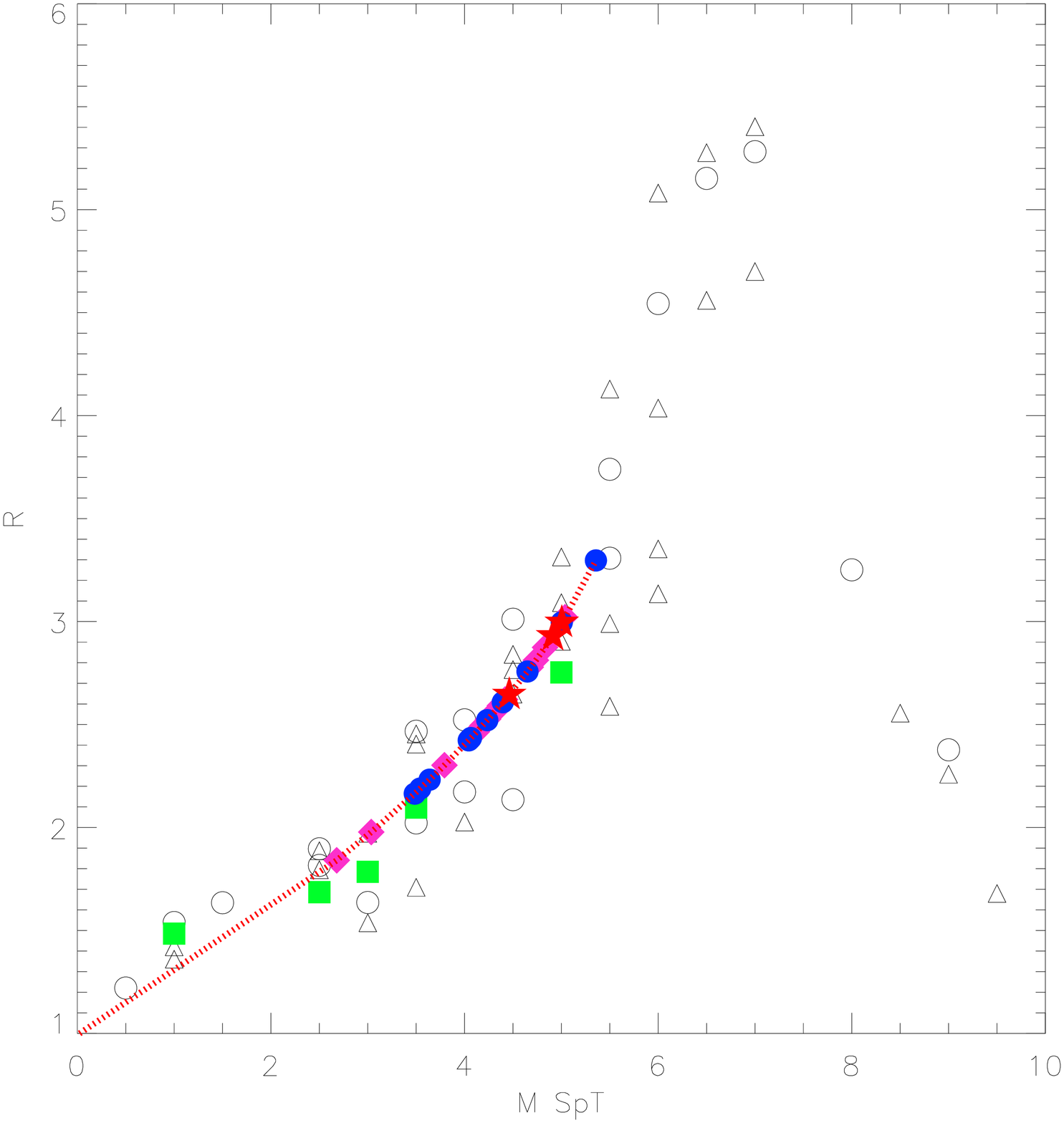}\\
	\includegraphics[width=0.49\textwidth]{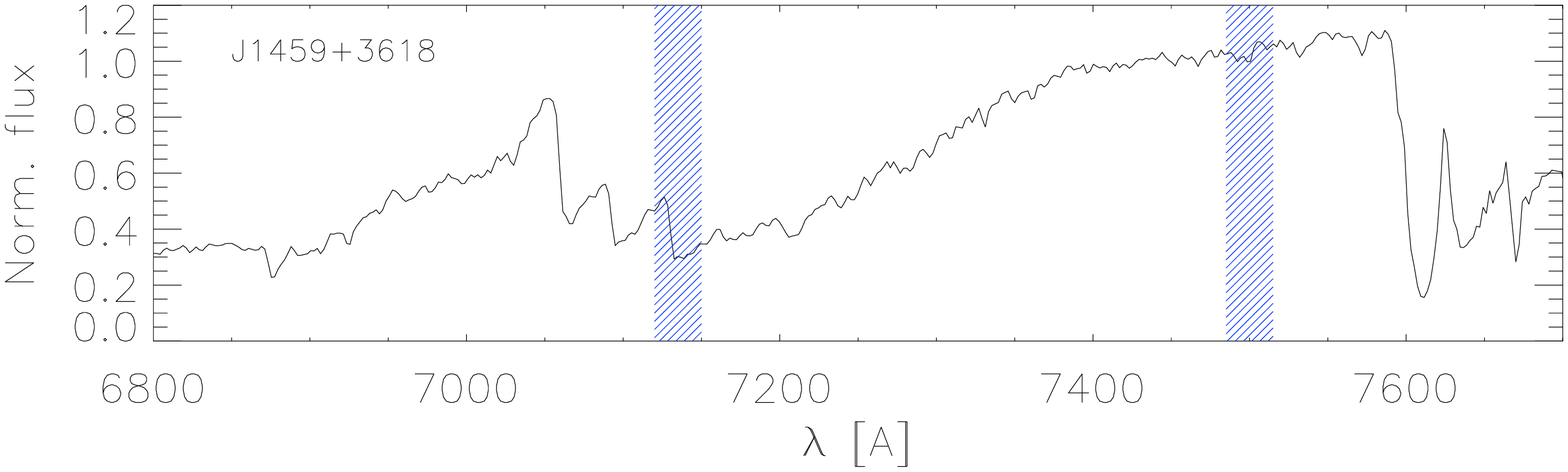}
	\caption{{\em Top panel}: our spectral index $\Re$ measured for M1--9\,V spectral type templates observed by \cite{2000ApJ...535..965L}  and
\cite{2002AJ....123.2828C} (black empty circles and triangles, respectively) compared with our 24 field M dwarfs (blue filled circles), 3 T~Tauri M-type objects (red filled stars), and 5 reference field dwarfs (green filled squares). 
{\em Bottom panel}: wavelength intervals used to define the Re index on an example spectrum.}
\label{figure.spectralindex}
\end{figure}

For the 27 stars, we derived our own spectral types with a custom-made spectral index.
Spectral indices have been extensively used in the classification of M dwarf spectra 
(\citealt{1991ApJS...77..417K}; 
\citealt{1995AJ....110.1838R}; 
\citealt{1996ApJ...469..706M, 1999AJ....118.2466M};
\citealt{2002AJ....123.3409H};
\citealt{2003AJ....125.1598L};
\citealt{2006AJ....131.3016S, 2006AJ....132.2665S}; 
\citealt{2011ApJ...727....6S}; 
\citealt{2011AN....332..821S}).

We took advantage of this knowledge for defining the $\Re$ index, which better fits the useful wavelength interval, resolution, and maximum efficiency of our IDS/INT spectra.
The numerator and denominator of the $\Re$ index are the fluxes contained in the 30 {\AA} bands centered on 7500 and 7135\,{\AA}, which correspond to the pseudo-continuum at the red side and to the minimum of the strong $\sim\!$7000--7350\,{\AA} TiO band, respectively.
Basically, $\Re$ is similar to the \cite{1996ApJ...469..706M} PC2 index (7560$\pm$20\,{\AA} / 7040$\pm$10\,{\AA}), which accounts mostly for the TiO and VO contributions, but with the numerator wavelength interval at the bottom of a deep water vapor band head. The  $\Re$ index is not sensitive to luminosity, $\log$ g or metallicity.
To minimize the dependence of our index on flux calibration calibration issues, we normalized our spectra to a pseudo-continuum traced by joining the highest points of the observed spectra (skipping H$\alpha$). 
The ratios were measured on the ``normalized'' spectra.
As templates for determining the $\Re$-spectral type relation, we used the spectra of 58 M1--9\,V stars of \cite{2000ApJ...535..965L} and \cite{2002AJ....123.2828C}, together with our 5 reference stars, normalized in the same way as the target spectra.
As illustrated by Figure.~\ref{figure.spectralindex}, the $\Re$ index has the advantage of having a fairly large dynamic range, covering values from about 1.0 to 5.5. 
We fit the $\Re$-SpT pairs to a parabola (i.e., ${\rm SpT}(\Re) = a + b \Re + c \Re^2$), and derived spectral types for our 27 targets with an estimated uncertainty of $\pm$0.5 spectral subtypes. The relation to estimate the spectral type, valid between M0.0\,V and M5.5\,V, was:

\begin{equation}
\small
SpT = -0.58\Re^2+4.8\Re-4.2
\end{equation}

The results are listed in the third column of Table \ref{table.additional}.
We calculate 18 new spectral types for the first time and improve previous determinations for other 7 dwarfs.
Derived spectral types vary from M2.5\,V in the case of J1132+1816 to M5.5\,V in the case of J2211+4059 (PM~I22489+1819), with the majority of them in the narrow interval from M3.5\,V to M4.5\,V, which fully backs our initial color criteria for selecting intermediate M dwarfs (the original cut in $r'-J$ was for selecting $>$M4\,V stars).
Some of the stars M4\,V and later are bright enough to be potential targets for exoplanet surveys, such as CARMENES \citep{2012SPIE.8446E..0RQ, 2013prpl.conf2K020C}.
Interestingly, the latest M dwarf in our sample had a spectral-type estimation from $V_{\rm phot}-J$ photometry at M7:\,V.
We agree with \citet{2013AJ....145..102L} and \citet{2013prpl.conf2K055M} that spectral types from this color are systematically later than those actually measured on real spectra (at least for spectral types later than M2--3\,V).

\subsection{Effective Temperatures and Surface Gravities}
\label{section.teff+logg}

\begin{figure}
   	\centering
    	\includegraphics[width=0.49\textwidth]{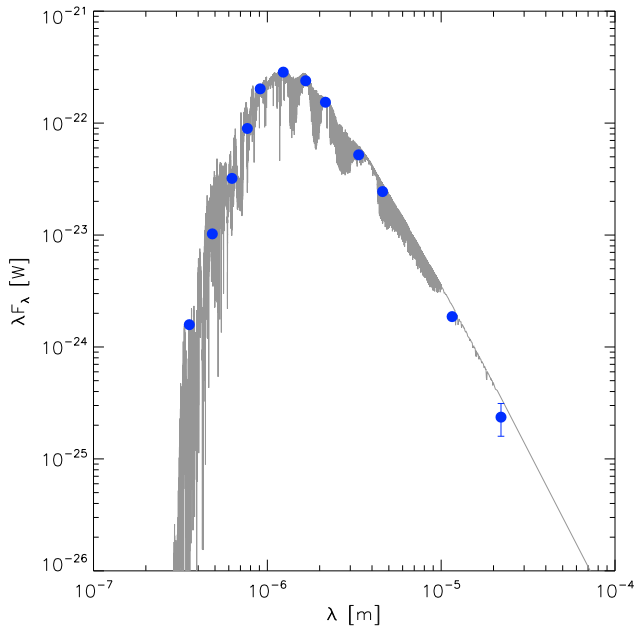}
 	\caption{Spectral energy distribution of J1459+3618 from 350\,nm ($u'$) to 22,100\,nm ($W4$), in blue filled circles, and the corresponding best-fit NextGen model provided by VOSA.}
\label{seds}
\end{figure}

We used another VO tool, the VO Spectral energy distribution Analyzer (VOSA\footnote{\tt http://svo2.cab.inta-csic.es/theory/vosa/}; \citealt{2008A&A...492..277B}), to derive effective temperatures ($T_{\rm eff}$) and surface gravities ($\log{g}$) of our 27 targets from fits of observed spectral energy distributions to theoretical models.
Apart from the CMC14 ($r'$) and 2MASS ($JHK_s$) photometric data, we also used those of the {\em Wide-field Infrared Survey Explorer} ($W$1--4; {\em WISE}, \citealt{2012yCat.2311....0C}) and, when available, the Fourth U.~S. Naval Observatory CCD Astrograph Catalogue ({\bf $Bg'Vi'$}; UCAC4, \citealt{2012yCat.1322....0Z}), and Sloan Digital Sky Survey ({\bf $u'g'$}; SDSS DR9, \citealt{2011ApJS..193...29A}).
We did not use the \citet{2011AJ....141..166H} $V$-band photometry of J1459+3618/RX~J1459.4+3618 because of an incorrect absolute calibration (besides, they found a period of photometric variability of 4.17\,d, but with a $V$-band amplitude of only 49\,mmag).
Key photometry of the 27 targets is provided in Table~\ref{table.photometry}.

In VOSA, we used the BT-Settl  theoretical models \citep{2014IAUS..299..271A} between 1\,600 and 4\,000\,K in $T_{\rm eff}$ and between 3.5 and 6.0 in $\log{g}$ for solar metallicity.
The uncertainty in the best fit was the size of the grid, which was of 100\,K in $T_{\rm eff}$ and 0.5 in $\log{g}$. Anayway, the VOSA $\log{g}$ values  have to be taken with caution and refined using other indicators.
 Figure.~\ref{seds} illustrates one of our VOSA fits as an example.

Derived values ranged between  2800 and 3400\,K in $T_{\rm eff}$, and 3.5 and 6.0 in $\log{g}$, which roughly match our spectral types and the surface gravity expected values for normal mid-M dwarfs in the field (e.g., \citealt{2013A&A...556A..15R}). 
As a matter of  fact, the star with the latest spectral type in our sample (J2211+4059, M5.5\,V) also had the lowest effective temperature ($T_{\rm eff}$ = 2900$\pm$100\,K).

We double-checked the VOSA values of $\log{g}$ with an atomic gravity-sensitive feature present in our spectra, the Na~{\sc i} doublet at 818.3--819.5\,nm (\citealt{1995MNRAS.272..630S}; \citealt{2006A&A...446..485G}; \citealt{2006AJ....131.3016S}; \citealt{2010A&A...517A..53M}; \citealt{2012AJ....143..114S}). 
At a given spectral type in low-mass stars (and brown dwarfs), the weaker the alkali doublet, the lower the gravity.
In its turn, low gravity is an indicator of youth (\citealt{1999ApJ...521..671B}; \citealt{2004ApJ...600.1020M}; \citealt{2005MNRAS.356.1583B}; \citealt{2013arXiv1311.7024S} and references therein).
We measured the pseudo-equivalent widths of the alkali doublet in our IDS/INT spectra with the IRAF task {\tt splot}.
The results, given in Table~\ref{table.additional}, showed that there are three M4.0--5.0\, stars with significantly weak sodium absorption, of pEW(Na~{\sc i}) $<$ 3\,{\AA}; two of them also had the lowest surface gravities ($\log{g}$ = 3.5$\pm$0.5) in the VOSA fits.
The other 24 stars had sodium absorptions typical of field dwarfs of the same spectral types (see Table~4 in \citealt{2012AJ....143..114S}).

The compiled photometry also allowed us to search for infrared excesses, which may be ascribed to circumstellar disks.
In particular, two stars, J0515+2336 and J0507+3730, had a significantly bright $W4$ magnitude (at 22.1\,$\mu$m) with respect to the other {\em WISE} and 2MASS magnitudes (Table~\ref{table.photometry}).
However, their apparent $W4$-band excess came instead from an incorrect background subtraction at very low Galactic latitudes (column $b$ in Table~\ref{table.basicdata}).
All in all, no star in our sample displayed a clear mid-infrared flux excess attributable to a disks.

\begin{table*}
%\renewcommand\tabcolsep{0.25pt}
%\caption{Photometry of the 27 investigated stars} 
%\label{table.photometry}
\caption{Photometry of the 27 investigated stars} 
\centering
\scriptsize
\begin{tabular}{lcccccc}
%\caption{Photometry of the 27 investigated stars} 
\hline
\hline
ID             & $u'$   &  $B$           &   $g'$     &    $V$          & $r'$        & $J$       \\
                 &[mag]   &   [mag]    &    [mag]    &  [mag]         & [mag]    & [mag]    \\ 
                 \hline
J0012+3028     & 17.884$\pm$0.013&  16.361$\pm$0.180     & 15.801$\pm$0.004       &  14.864$\pm$0.060          & 14.216        & 10.242$\pm$0.023 \\
J0013+2733     & 19.224$\pm$   0.039 &  16.753$\pm$0.080     & 16.012$\pm$0.003       &   15.115$\pm$0.060     & 14.437        & 10.431$\pm$0.020\\  
J0024+2626     & 18.122$\pm$0.018     &  16.343$\pm$0.130     & 15.550$\pm$0.003       & 14.735$\pm$0.040       & 14.131        & 10.222$\pm$0.020\\
J0058+3919     & 17.552$\pm$0.01        &   15.837$\pm$0.050    &  15.078$\pm$0.003      &  14.168$\pm$0.030      & 13.598       &  9.561$\pm$0.026  \\
J0122+2209     & ...                                     &   14.722$\pm$0.030    &  13.763$\pm$0.04        &  12.996$\pm$0.030      & 12.375        &  8.412$\pm$0.021  \\
 J0156+3033      & ...                                    &16.532$\pm$0.050       &  15.786$\pm$0.08        &  15.105$\pm$0.050      & 14.449        & 10.323$\pm$0.023  \\
 J0304+2203       & ...                                   & 17.164$\pm$0.030      &   16.199$\pm$0.05       &  15.564$\pm$0.030      & 14.931        & 10.486$\pm$0.022\\
 J0326+3929$^{b}$&17.991$\pm$0.013&  ...                                     &  15.546$\pm$0.003     & ...                                      & 13.968         &  9.998$\pm$0.025\\
 J0327+2212    & ...                                      & 16.252$\pm$0.120      & 15.314$\pm$0.04         &14.654$\pm$0.090        & 14.055         & 10.044$\pm$0.022\\   
 J0341+1824     & ...                                     &  16.577$\pm$0.030     & 15.669$\pm$0.03         &14.958$\pm$0.100        & 14.412         & 10.484$\pm$0.021  \\
 J0342+2326     & ...                                     &  16.609$\pm$0.060     &  15.667$\pm$0.05        &14.984$\pm$0.060        & 14.267        & 10.202$\pm$0.022\\
 J0422+2439$^{b}$&18.426$\pm$0.014&  ...                                     &  16.092$\pm$0.003     &...                                       & 14.529        &  9.648$\pm$0.021 \\
 J0424+3706    &17.711$\pm$0.012        & 16.337$\pm$0.020      & 15.718$\pm$0.005      &14.668$\pm$0.030        & 14.203       & 10.191$\pm$0.026 \\
J0435+2523$^{b}$& 18.303$\pm$0.014& ...                                     &16.189$\pm$0.004       &...                                       & 14.669        & 10.272$\pm$0.022   \\
J0439+2333     &18.433$\pm$0.015       & 17.028$\pm$0.120      & 16.240$\pm$ 0.004     &15.392$\pm$0.010        & 14.856        & 10.479$\pm$0.023 \\
 J0507+3730     &...                                      & 16.986$\pm$0.230      & 16.018$\pm$0.03         &15.349$\pm$0.060        & 14.713        & 10.284$\pm$0.020\\
 J0515+2336     &...                                      &  16.977$\pm$0.010     & 16.031$\pm$0.05         &15.245$\pm$0.040       & 14.574        & 10.186$\pm$0.030   \\
 J0630+3003      &...                                     & ...                                      &...                                      &...                                        & 14.020       & 10.045$\pm$0.018   \\
 J0909+2247     &18.856$\pm$0.018       & 17.172$\pm$0.010      &  16.230$\pm$0.004     &15.266$\pm$0.010        & 14.656       & 10.474$\pm$0.023\\
 J1132+1816     &18.223$\pm$0.016       &  16.354$\pm$0.070     &  15.625$\pm$0.004     &14.751$\pm$0.050        & 14.137        & 10.175$\pm$0.023 \\
 J1241+1905$^{b}$&18.369$\pm$0.017&  16.750$\pm$0.080      &   15.882$\pm$0.004    &15.048$\pm$0.030       & 14.284        & 10.368$\pm$0.022 \\
 J1459+3618     &18.183$\pm$0.012       &16.566 $\pm$0.040      & 15.784$\pm$0.004       &14.845$\pm$0.080       & 14.292         & 10.257$\pm$0.018 \\
J1518+2036     &...                                       &  ...                                    &  ...                                     &15.455$\pm$0.010        & 14.786       & 10.119$\pm$0.021 \\
  J1574+2241     &17.438$\pm$0.013        & 15.955$\pm$0.040      &  15.139$\pm$0.004     &14.197$\pm$0.010       & 13.589         &  9.543$\pm$0.022 \\   
 J2211+4059     &18.465$\pm$0.015        & 16.963$\pm$0.090      &  16.081$\pm$0.003     &15.139$\pm$0.010       & 14.371        &  9.725$\pm$0.020   \\
 J2248+1819    &17.838$\pm$0.011         & 16.115$\pm$0.080      &   15.461$\pm$0.005    &14.535$\pm$0.040       & 13.886       &  9.957$\pm$0.021  \\
 J2259+3736      &...                                     &   16.942$\pm$0.020     &  16.043$\pm$0.05        &15.357$\pm$0.120        &14.638     & 10.378$\pm$0.029   \\
 \hline
 ID            & $H$         & $K_{\rm s}$           &  $W1$                         &  $W2$                         & $W3$                          & $W4$                  \\
                  &[mag]   &   [mag]                       &    [mag]                         &  [mag]                         & [mag]                          & [mag]                     \\ 
\hline    
 J0012+3028        & 9.683$\pm$0.022     & 9.410$\pm$0.021     & 9.233$\pm$0.022        &9.039$\pm$0.020        &8.929$\pm$0.026        &8.784$\pm$0.386 \\
 J0013+2733        & 9.837$\pm$0.020      & 9.581$\pm$0.018    & 9.376$\pm$0.024        &9.210$\pm$0.022        &9.077$\pm$0.027        &8.592$\pm$0.317 \\
 J0024+2626        & 9.592$\pm$0.019       & 9.299$\pm$0.017    & 9.188$\pm$0.023       &8.959$\pm$0.020       &8.819$\pm$0.025         &8.543$\pm$0.298 \\
 J0058+3919         & 8.947$\pm$0.029       & 8.680$\pm$0.018     & 8.489$\pm$0.023        &8.303$\pm$0.020        &8.150$\pm$0.017        &8.009$\pm$0.133 \\
 J0122+2209         & 7.820$\pm$0.016       & 7.537$\pm$0.017    & 7.341$\pm$0.027        &7.167$\pm$0.020        &7.055$\pm$0.016        &6.856$\pm$0.060 \\
 J0156+3033        & 9.718$\pm$0.031       & 9.449$\pm$0.021    & 9.267$\pm$0.023       & 9.074$\pm$0.019       & 8.925$\pm$0.028       & 9.075$\pm$0.540 \\
 J0304+2203        & 9.933$\pm$0.021       & 9.655$\pm$0.018    & 9.427$\pm$0.023       & 9.226$\pm$0.022       & 9.098$\pm$0.032       & 8.476: $^{a}$         \\
 J0326+3929$^{b}$& 9.412$\pm$0.026       & 9.084$\pm$0.017    & 8.884$\pm$0.028       & 8.701$\pm$0.026       & 8.616$\pm$0.030       & 8.017$\pm$0.274       \\
 J0327+2212       & 9.477$\pm$0.020     & 9.194$\pm$0.017    & 9.077$\pm$0.022       & 8.872$\pm$0.020       & 8.730$\pm$0.029       & 8.812$\pm$0.434       \\
 J0341+1824       & 9.858$\pm$0.023       & 9.643$\pm$0.023   & 9.435$\pm$0.023       & 9.245$\pm$0.021       & 9.175$\pm$0.036       & 8.866:$^{a}$ \\
 J0342+2326      & 9.545$\pm$0.023       & 9.316$\pm$0.023    & 9.158$\pm$0.025       & 8.979$\pm$0.022       & 8.817$\pm$0.028       & 8.076$\pm$0.354       \\
 J0422+2439$^{b}$& 8.947$\pm$0.022       & 8.651$\pm$0.020     & 8.467$\pm$0.032       & 8.185$\pm$0.030       & 8.162$\pm$0.028       & 8.014$\pm$0.269       \\
 J0424+3706     & 9.553$\pm$0.030       & 9.340$\pm$0.020    & 9.184$\pm$0.023       & 9.009$\pm$0.019       & 8.830$\pm$0.030       & 8.193:$^{a}$  \\
 J0435+2523$^{b}$& 9.618$\pm$0.030      & 9.331$\pm$0.021    & 9.216$\pm$0.023       & 8.997$\pm$0.020       & 8.821$\pm$0.027       & 8.249$\pm$0.262       \\
 J0439+2333      & 9.893$\pm$0.021       & 9.617$\pm$0.017    & 9.471$\pm$0.024       & 9.258$\pm$0.020       & 9.131$\pm$0.033       & 8.530$\pm$0.381       \\
 J0507+3730     & 9.703$\pm$0.021       & 9.397$\pm$0.018    & 9.223$\pm$0.022       & 9.008$\pm$0.019       & 8.686$\pm$0.024       & 7.656$\pm$0.133       \\
 J0515+2336     & 9.602$\pm$0.035       & 9.306$\pm$0.024     & 9.067$\pm$0.024       & 8.897$\pm$0.020       & 8.672$\pm$0.026       & 7.397$\pm$0.175       \\
 J0630+3003      & 9.484$\pm$0.018       & 9.208$\pm$0.018     & 9.025$\pm$0.023       & 8.847$\pm$0.018       & 8.700$\pm$0.029       & 8.090$\pm$0.256       \\
 J0909+2247    & 9.915$\pm$0.032       & 9.616$\pm$0.018      & 9.367$\pm$0.021       &9.206$\pm$0.020        & 8.996$\pm$0.030       &8.470:$^{a}$           \\
 J1132+1816     & 9.599$\pm$0.030      & 9.338$\pm$0.022      &9.138$\pm$0.024        &8.961$\pm$0.021        & 8.821$\pm$0.024       &9.182$\pm$0.534\\
 J1241+1905$^{b}$& 9.792$\pm$0.026       & 9.477$\pm$0.018      & 9.314$\pm$0.023       &9.136$\pm$0.019        & 8.978$\pm$0.025       &8.336$\pm$0.225\\ 
 J1459+3618     & 9.647$\pm$0.016       & 9.377$\pm$0.016    & 9.230$\pm$0.022       & 9.070$\pm$0.020       & 8.939$\pm$0.021       &9.123$\pm$0.353 \\
 J1518+2036     & 9.606$\pm$0.022       & 9.268$\pm$0.019    & 9.039$\pm$0.022       & 8.839$\pm$0.020       & 8.641$\pm$0.019       & 8.515$\pm$0.194       \\
 J1574+2241     & 8.932$\pm$0.030      & 8.647$\pm$0.022     & 8.475$\pm$0.023       & 8.305$\pm$0.020       & 8.145$\pm$0.017       & 8.007$\pm$0.157       \\
 J2211+4059     & 9.097$\pm$0.017       & 8.790$\pm$0.016     &8.565$\pm$0.021        & 8.404$\pm$0.020       & 8.213$\pm$0.019       &8.661$\pm$0.304 \\
 J2248+1819    & 9.388$\pm$0.020       & 9.119$\pm$0.017      &8.945$\pm$0.022       & 8.760$\pm$0.020     & 8.604$\pm$0.023       &8.523:$^{a}$           \\
 J2259+3736    & 9.890$\pm$0.037      & 9.535$\pm$0.024      &0.257$\pm$0.023       & 9.062$\pm$0.020      & 8.925$\pm$0.025       &8.509$\pm$0.264\\
\hline
\end{tabular}
%\caption{Photometry of the 27 investigated stars} 
\label{table.photometry}
 \begin{minipage}{15cm} 
$^{a}${Poor quality flags.}\\
$^{b}${The SDSS $i'$ band was not use to estimated the $T_{\rm eff}$}
\end{minipage}
%\end{sidewaystable*}
\end{table*}
%\end{landscape}

\subsection{Activity}

\begin{figure}
   	\centering
 	\includegraphics[width=0.5\textwidth]{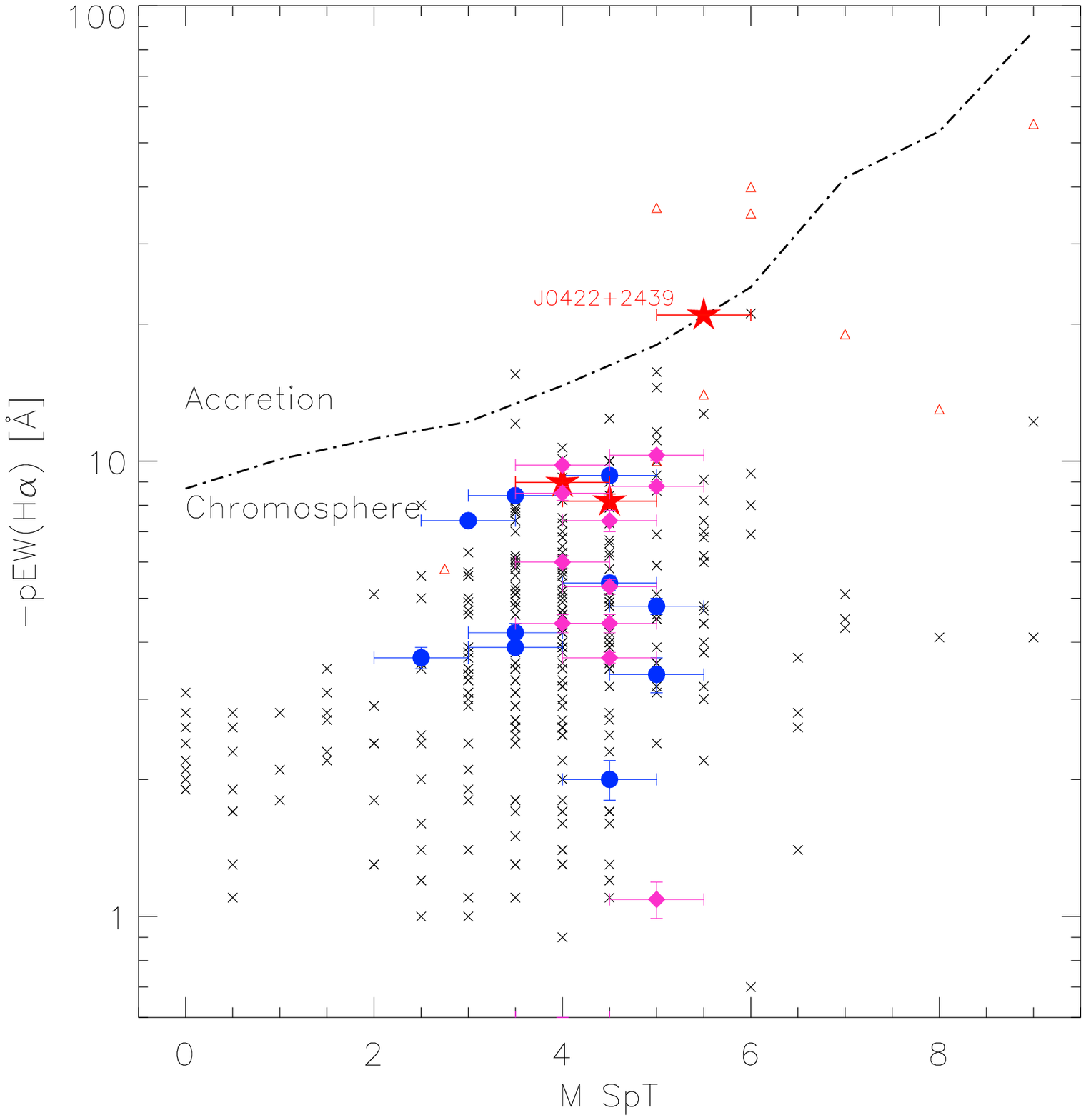}
  	\caption{Pseudo-equivalent widths of H$\alpha$ as a function of spectral type with the \cite{2003AJ....126.2997B} accretion-chromosphere boundary.
Small (black) crosses are field M dwarfs from \cite{1999A&AS..139..555G}, (red) triangles are confirmed Taurus-Auriga members from  \cite{2001ApJ...561L.195M}, \cite{2003ApJ...590..348L}, and \cite{2003ApJ...592..266M}, (magenta and blue) filled circles are the 24 (new and known) field stars investigated here, and (red) filled stars are our three new Taurus-Auriga member candidates, including J0422+2439.}
\label{actividad}
\end{figure}

We tried to quantify the magnetic activity of the stars in our sample.
First, we measured pseudo-equivalent widths of the H$\alpha$ $\lambda$656.3\,nm line, pEW(H$\alpha$)s, in our IDS/INT spectra.
Error bars for each target were assigned by manual repetition of measurements making educated visual inspections of the continuum levels and the line limits.
As expected for intermediate- and late-type M dwarfs (\citealt{1996AJ....112.2799H}; \citealt{2000AJ....120.1085G}; \citealt{2004AJ....128..426W}), most of our stars showed H$\alpha$ in emission.
Indeed, two of the three non-H$\alpha$ emitters are the earliest stars in our sample (M2.5--3.0\,V).
However, one of the stars displayed an H$\alpha$ pseudo-equivalent width that stood out among the other measurements: J0422+2439, with pEW(H$\alpha$) = --20.9$\pm$0.5\,{\AA}.
We used the \cite{2003AJ....126.2997B} empirical criterion for ascertaining the origin of the H$\alpha$ emission.
As illustrated by Figure.~\ref{actividad}, the emission of all stars in our sample except J0422+2439 is consistent with chromospheric activity. J0422+2439, one of the three low-gravity stars described in Section~\ref{section.teff+logg}, showed H$\alpha$ emission very close to the criterion boundary separating accretion and chromospheric emission.

There were two stars, J0122+2209/G~34--53 and J1459+3618/RX~J1459.4+3618, for which H$\alpha$ emission had been investigated previously by \cite{2002AJ....124.2868M} (Table~\ref{table.additional}).
Their and our measurements of pEW(H$\alpha$) match each other within the uncertainties.

Second, we searched for counterparts in the {\em ROSAT} All-Sky Bright and Faint Survey Catalogues \citep{1999A&A...349..389V}.
We applied a search radius of 30\,arcsec due to the low {\em ROSAT} astrometric precision. 
Of the 27 stars in our sample, 11 stars ($\sim$40\,\%) had appreciable emission in the 0.2--2.0\,keV energy band at the time of the {\em ROSAT} observations.

The brightest star in our sample in the visible and near-infrared, J0122+2209/G~34--53, also has  the highest X-ray count rate by far.
Perhaps because of that reason, it has been the subject of a few all-sky X-ray surveys for low-mass stars (\citealt{1998ApJ...504..461F}; \citealt{2003A&A...406..535Z}; \citealt{2003A&A...403..247F}; \citealt{2009ApJS..184..138H}).
Another three known dwarfs had also been cataloged  as X-ray emitters: J0156+3033/NLTT~6896 \citep{2012Obs...132....1C}, J1459+3618/RX J1459.4+3618 (\citealt{1998ApJ...504..461F}), and J2211+4059/1RXS J221124.3+410000 (\citealt{2009ApJS..184..138H}).
In this work, we report for the first time the X-ray emission of one known M dwarf, J0058+3919/PM I00580+3919, and six new M dwarfs.

For the 11 X-ray stars, we calculated the distance-independent parameter $F_X / F_J$ ($\equiv L_X / L_J$), which is a proxy of $L_X / L_{\rm bol}$ (\citealt{2010A&A...521A..45C}).
We computed $F_X$ from the X-ray count rates and hardness ratios as in \citet{1995ApJ...450..392S}.
Six stars had $F_X / F_J$ ratios above $0.7\times10^{-3}$, with only one of them having a ratio of $\sim1.1\times10^{-3}$.
This relatively strong X-ray emitter is just J2211+4059/1RXS J221124.3+410000, the primary of an ultra fragile binary system.
The other five intense emitters are
J0122+2209/G~34--53 (the brightest star of our list),
J0156+3033/NLTT~6896 (the primary of the Koenigstuhl~4 wide binary system),
J0422+2439 (the accreting star with strong H$\alpha$ emission and low surface gravity), and
J0304+2203 and
J0507+3730  (two new, M4.5-5.0\,V stars identified in this work).
The latter three intense X-ray emitters are shown here for the first time.
The other six stars with {\em ROSAT} data are relatively faint X-ray emitters.

\subsection{Three New Member Candidates in Taurus}
\label{section.taurus}

\begin{figure*}
 	\centering
	\includegraphics[width=0.32\textwidth]{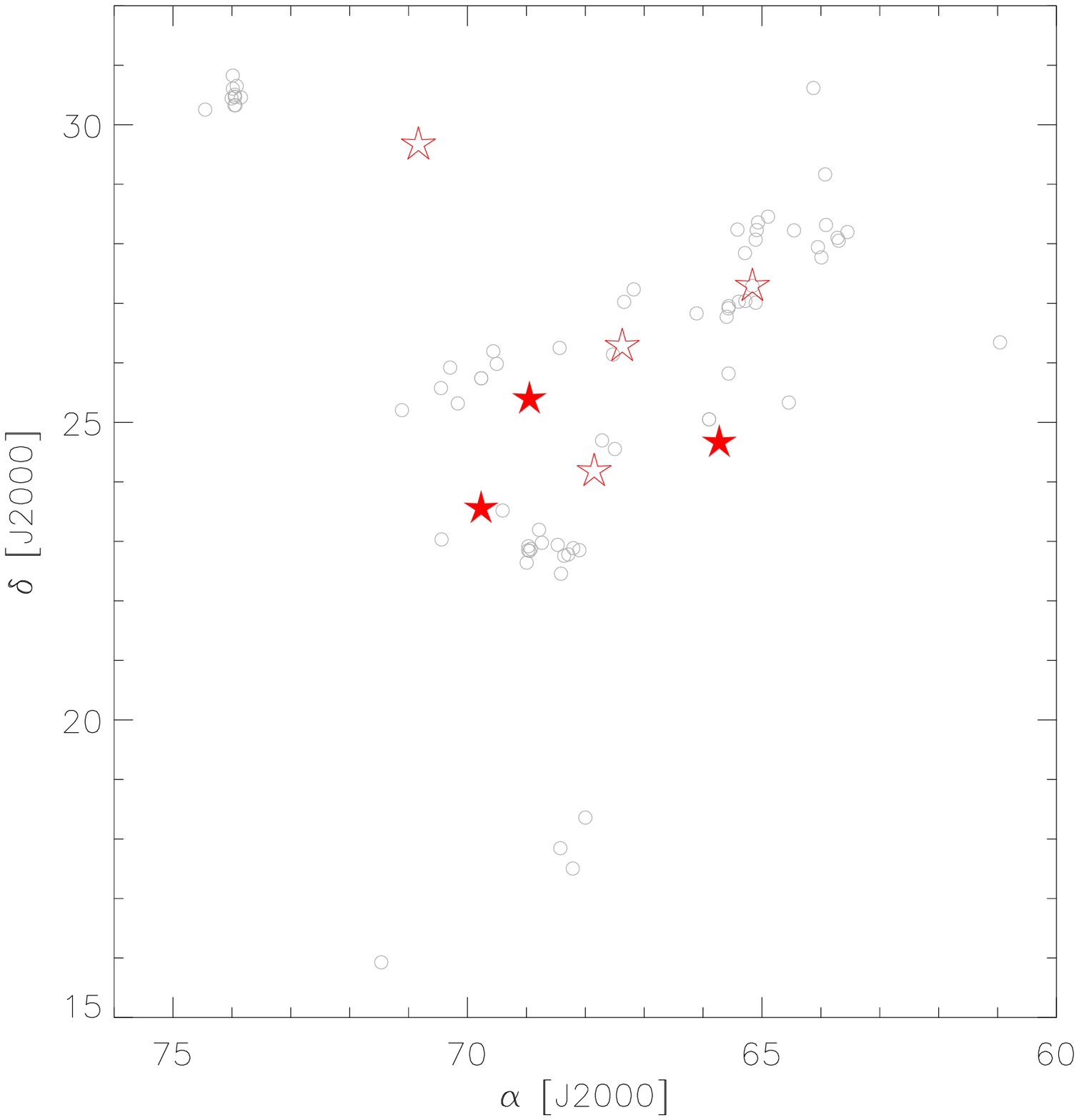}
  	\includegraphics[width=0.32\textwidth]{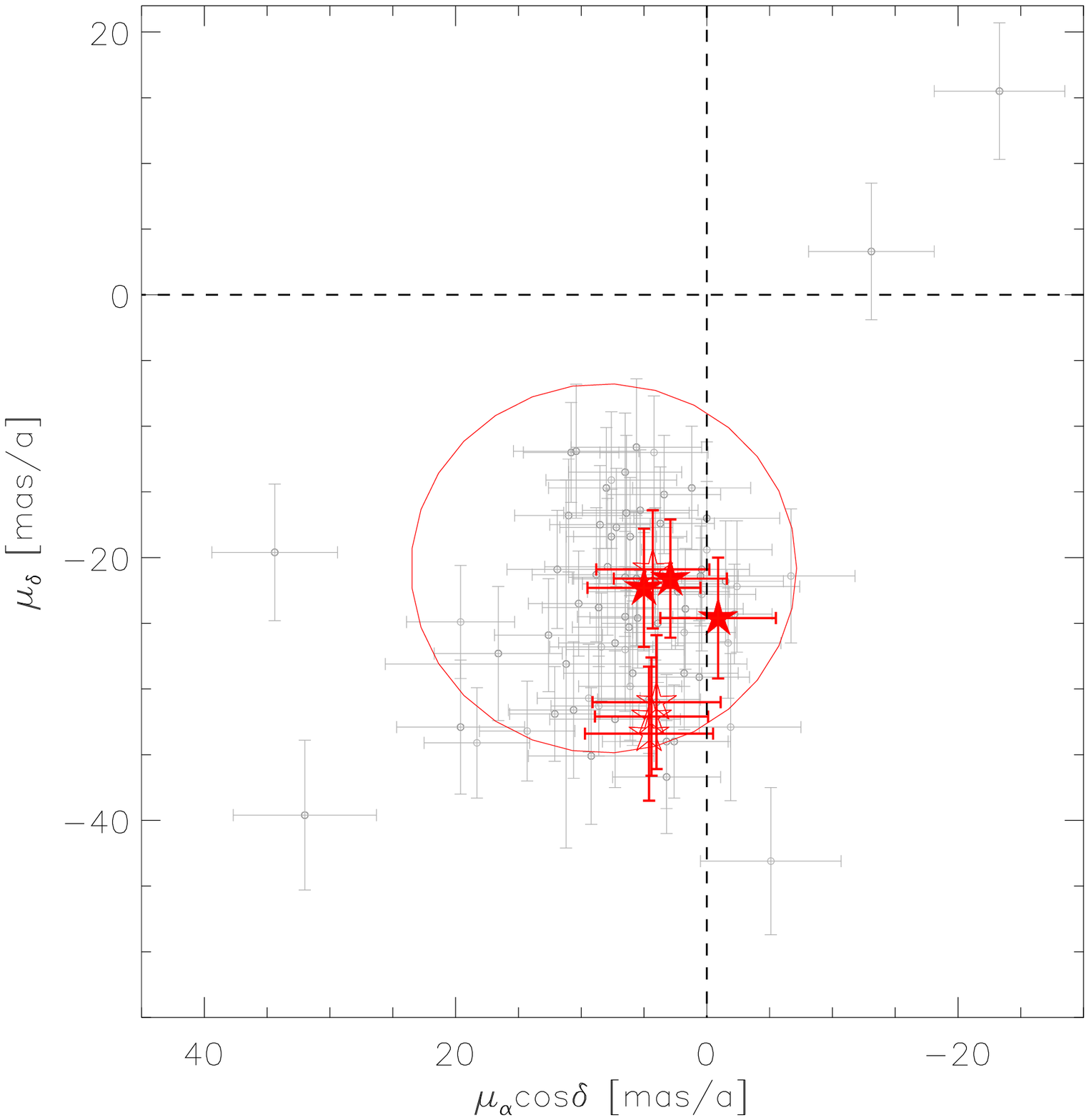} 
 	\includegraphics[width=0.32\textwidth]{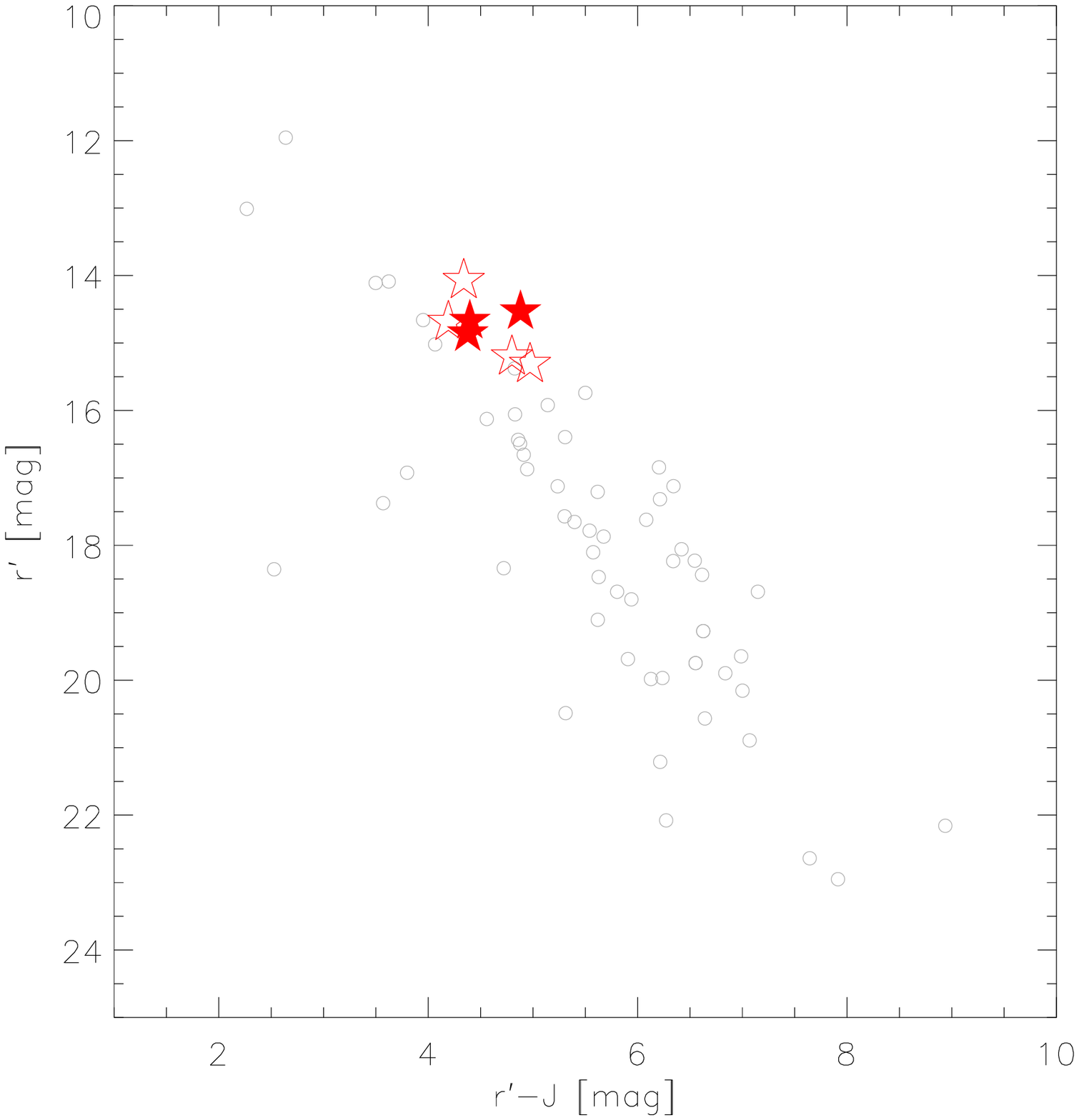}
	\caption{
{\em Left panel:} spatial distribution of candidate members in the Taurus-Auriga star-forming region.
(Red) filled stars are our three new candidates with IDS/INT spectroscopy, (red) open stars are the four known Taurus-Auriga T~Tauri stars identified in the 2MASS-CMC14 cross-match and listed in Table~\ref{table.known}, and (gray) open circles are Taurus-Auriga members from  \cite{2004ApJ...617.1216L}, \cite{2006ApJ...647.1180L, 2009ApJ...703..399L}, and \cite{2006A&A...446..485G}.
{\em Middle panel: } same as the left panel, but for the proper-motion diagram. 
The ellipse indicates the average values and standard deviations of proper motions in Taurus-Auriga from \citet{2006A&A...460..499B}, 
{\em Right panel: } same as the left panel, but for the $r'$ vs. $r'-J$ color-magnitude diagram.}
\label{figure.taurus}
\end{figure*}

The strong H$\alpha$ emission, low surface gravity, and relatively intense X-ray emission of the possibly accreting star J0422+2439 led us to investigate it in detail.
Its coordinates, as well as those of the other two stars with weak Na~{\sc i} absorption, resembled those of the four known young T~Tauri stars in Table~\ref{table.known}, so we considered their membership in the young star-forming region of Taurus-Auriga ($\tau \sim$ 1--2\,Ma, $d \sim$ 140\,pc; \citealt{2008hsf1.book..405K}).

First of all, we compiled an exhaustive list of Taurus-Auriga members and member candidates from \cite{2004ApJ...617.1216L}, \cite{2006ApJ...647.1180L, 2009ApJ...703..399L}, and \cite{2006A&A...446..485G} and cross-matched that list with 2MASS, CMC14, and PPMXL.
Figure.~\ref{figure.taurus} illustrates our analyses. From left to right, our three new Taurus-Auriga member candidates with spectroscopy (and the four T~Tauri stars in Table~\ref{table.known}) 
(1) are spatially located towards the densest filaments of Taurus-Auriga (see also Figure 1 in \citealt{2009ApJ...703..399L}),
(2) have proper motions consistent with membership in Taurus-Auriga (\citealt{2006A&A...460..499B}; \citealt{2013ApJ...771..110M}), 
and (3) follow the Taurus-Auriga sequence in optical-near-infrared color-magnitude diagrams\footnote{Note the previously known Taurus-Auriga candidates with discordant proper motions or blue $r'-J$ colors for their $r'$ magnitudes (CFHT-BD-Tau 19, KPNO-Tau 6, 2MASS J04201611+2821325, 2MASS J04202144+2813491, ITG\,34 and FS\,115. }.
Taking into account these facts, the low surface gravity of the three new member candidates, and the intense H$\alpha$ and X-ray emission of the accreting star, we concluded that J0422+2439, J0435+2523, and J0439+2333 do belong to Taurus-Auriga.

We further characterized the three identified T~Tauri stars.
At $d \sim$ 140\,pc, the Taurus-Auriga distance modulus is $m-M \sim$ 5.7\,mag.
With the $J$-band apparent magnitudes, the maximum Êabsolute magnitudes range between 3.9 and 4.7\,mag.
Actual absolute magnitudes must be brighter than that interval because of variable extinction toward Taurus-Auriga.
The detection of J0422+2439 by {\em ROSAT} implies that it is located in the closer side of the cloud, with subsequent low extinction (X-rays are absorbed by interstellar dust and gas).
Something similar happens to J0435+2523 and J0439+2333, whose SEDs and colors do not deviate significantly from the other M dwarfs in the field.
By assuming conservatively that the $J$-band extinction is lower than 0.5\,mag and using the NextGen models (Baraffe et~al. 1998) at 1--2\,Ma, we derived theoretical masses in the interval 0.17--0.57\,$M_\odot$ for the three new T~Tauri low-mass stars, well above the hydrogen burning mass limit.
The three young stars could be at distances slightly closer than 140\,pc because they are on the near sides of the clouds, which are rather large.
We will know their distances with the advent of the ESA/{\em Gaia}/ space mission.

Some authors have argued against the universality of the initial mass function based on the hypothetical deficit of low-mass stellar objects ($<$0.5\,$M_{\sun}$, M2--4) in Taurus-Auriga (\citealt{1998AJ....115.2074B,2002ApJ...580..317B}; \citealt{2003ApJ...590..348L,2009ApJ...703..399L}; \citealt{2006A&A...446..485G}; \citealt{2007A&A...468..353G}; \citealt{2007A&A...468..405S}; \citealt{2013MmSAI..84..948P}). The detection of three stars just in the deficit range may point out that the initial mass function in Taurus-Auriga is actually standard, and that previous surveys had not been successful enough for detecting intermediate M dwarfs. Simple VO surveys such as the one presented here may cover that gap.

\subsection{Distances}
\label{section.distance}

The three young stars in Taurus-Auriga are consequently located at $d \sim$ 140\,pc (\citealt{2008hsf1.book..405K} and references therein).
However, some of our targets were expected to be located very close to the Sun because of their late spectral types and relative brightness.
We used the absolute magnitude $M_J$-spectral type relation in \cite{2008A&A...488..181C} for deriving spectro-photometric distances for the 24 investigated field M dwarfs, which are given in Table~\ref{table.additional}.
We assumed generous uncertainties in the $M_J$-SpT relation, apart from those in our SpT determination and the 2MASS $J$ magnitudes, which translated into typical error bars of about 20\,\%.
There have been previous determinations of the distances to three known dwarfs based solely on $VI$-band (\citealt{1998ApJ...504..461F}) and optical and near-infrared photometry \citep{2012Obs...132....1C}.
Those determinations are consistent with ours within conservative error bars.

All of our field M dwarfs except four (which are M2.5--3.5\,V stars) are located at less 25\,pc.
Of them, seven are at 15\,pc or less, of which only three were previously known (J0122+2209/G~34--53 at 8$\pm$2\,pc, J0156+3033/NLTT~6496 at 15$\pm$4\,pc, and J2211+4059/1RXS J221124.3+410000 at 9$\pm$2\,pc). 
The other four were identified in this work for the first time (J0012+3028, J0507+3730, J1518+2036, and J2259+3738 at 12--15\,pc).

\subsection{Common Proper Motion Pairs}

\begin{table}
%\scripsize
\centering
\renewcommand\tabcolsep{2.5pt}
\caption{Relative Astrometry of the LSPM J0326+3929EW (Koenigstuhl~7\,AB) Common Proper Motion Pair} 
\label{table.binary} 
\begin{tabular}{cc c l}
\hline\hline
\noalign{\smallskip}
$\rho$ 			& $\theta$	         		& Date		& Origin		\\ 	 
{[arcsec]} 			& [deg]               		&                          &            		\\	   	             
\noalign{\smallskip}
\hline
\noalign{\smallskip}
6.5$\pm$0.5 		& 236$\pm$6		& 1955 Feb 13	& POSS-I Red 	\\         
5.7$\pm$0.5 		& 229$\pm$6		& 1989 Sep 29	& POSS-II Red 	\\         
6.2$\pm$0.5 		& 229$\pm$6		& 1994 Nov 28	& POSS-II Blue \\         
5.9$\pm$0.5 		& 228$\pm$6    	& 1995 Nov 14	& POSS-II IR 	\\         
6.37$\pm$0.06      	& 227.7$\pm$1.0    	& 1998 Nov 1 & 2MASS  	\\         
6.38$\pm$0.10      	& 227.3$\pm$1.0    	& 2001 Dec 26 & CMC14		\\         
6.45$\pm$0.09      	& 227.4$\pm$1.0    	& 2003 Jan 7 	& SDSS-DR9	\\
6.70$\pm$0.11      	& 229.4$\pm$1.0    	& 2010 Jul 1 	& {\em WISE}	\\
\noalign{\smallskip}
\hline
\end{tabular}
\end{table}

We took advantage of the Aladin sky atlas and the VOTable Plotting tool VOplot to look for proper-motion companions to our 27 stars.
We loaded PPMXL data in a circular area of 30\,arcmin radius centered on our targets and plotted proper-motion diagrams ($\mu_{\delta}$ versus $\mu_{\alpha} \cos{\delta}$).
We recovered one known binary system (K\"onigstuhl 4~AB; \citealt{2012Obs...132....1C}) and reported and characterized for the first time another one.

The new binary, not tabulated in the Washington Double Star Catalog (\citealt{2001AJ....122.3466M}), consists of J0326+3929/LSPM J0326+3929E and its close companion \object{LSPM~J0326+3929W}.
The two stars were reported first by \citet{2005AJ....129.1483L}, who did not provide any clues of their possible binarity (the \citealt{2005AJ....129.1483L} catalog is a very useful source of new proper motion pairs identified by amateur astronomers -- e.g., \citealt{2012JDSO....8...73L}; \citealt{2013JDSO....8...260R}).
However, because of its angular separation that is  shorter than 55\,arcsec and identical proper motions, the pair probably is one of the 19,836 highly probable wide binaries reported by  \citet{2011ASPC..448.1375L}.

We applied the method of \citep{2007ApJ...667..520C} of confirming membership in a common proper motion pair by comparing multi-band photometry (in this case, $u'g'r'i' JHK_s W1 W2 W3$) of the two components and measuring constant angular separation $\rho$ and position angle $\theta$  on a long time baseline.
Table~\ref{table.binary} summarizes our astrometric analysis of SuperCOSMOS digitizations of the First and Second Palomar Observatory Sky Survey (\citealt{2001MNRAS.326.1279H}) and other public databases.
With a significant proper motion of $\mu \sim$ 160\,mas\,a$^{-1}$, the two stars would be separated by up to 15\,arcsec in the 1955 POSS-I images if the secondary star were in the background.
However, $\rho$ and $\theta$ kept constant at 6.3$\pm$0.3\,arcsec and 229$\pm$3\,deg in a 55.4\,a long interval, from which we concluded that the two stars travel together. Both LSPM J0326+3929E and W were saturated in all POSS photographic plates and, because of their proximity, there were large uncertainties in the determination of the photocentroids.
We determined more precise mean angular separation and position angle at $\rho$ = 6.48$\pm$0.16\,arcsec and $\theta$ = 227.9$\pm$1.0\,deg by averaging only the last four astrometric epochs ($\Delta t$ = 11.7\,a).
With the distance computed in Section~\ref{section.distance} ($d$ = 17$\pm$4\,pc), we derived a projected physical separation of $s$ = 110$\pm$30\,au. %$s = \rho d$
From the spectral type of the primary, M4.5\,V, and the magnitude differences between the two components, of $\Delta r'$ = 0.357$\pm$0.004\,mag and $\Delta J$ = 0.28$\pm$0.04\,mag, we estimated a spectral type M5.0:\,V for the secondary.

\section{Discussion and conclusions}
\label{section.discussion}

We showed the potential of the Virtual Observatory for finding new bright nearby M dwarfs, some of which can be targeted by current or forthcoming exoplanet surveys.
In this pilot program, we cross-matched the photometric CMC14 ($r'$) and 2MASS ($JHK_s$) catalogs in the whole overlapping area of 25\,078\,deg$^2$, imposed color restrictions in $r'-J$ and $J-K_s$, and selected 828 sources brighter than $J$ = 10.5\,mag for follow-up.  
Some of them turned out to be background M giants or even reddened, massive early-type stars in distant open clusters. Proper motions were used in a second step in giving priorities in the spectroscopic follow-up.

We used the Intermediate Dispersion Spectrograph at the 2.5\,m Isaac Newton Telescope for obtaining low-resolution optical spectroscopy of 27 targets, 25 of which had not been spectroscopically analyzed before. 
We determined spectral types with a custom-made spectral index, $\Re$, which accounts mostly for the absorption of a TiO band at 7100--7500\,\AA.
Derived spectral types of all the stars ranged between M2.5\,V and M5.5\,V, with the bulk of them in the narrower M3.5--5.0\,V interval, which demonstrated the success of our search.
 
In spite of their relative brightness, $J <$ 10.2\,mag in seven cases (and $J <$ 9.7\,mag in one case), 16 (60\,\%) M dwarfs had escaped previous surveys and are, therefore, discovered and characterized here for the first time.
This fact may be due to that our survey, contrary to most searches for M dwarfs, was purely photometric and that most of our stars fell above the main-sequence locus in a reduced-proper-motion diagram. 
That is, proper motions of our targets are lower than average for typical dwarfs of the same $r'-J$ colors.
Without an appropriate radial-velocity study, one cannot deduce low total Galactic velocity from low tangential velocity (M dwarfs can move fast in the visual direction instead), but one can at least conclude that proper-motion surveys are inefficient in the identification of slow M dwarfs, even if they are bright and nearby.

Indeed, among our 27 M dwarfs, there are two stars at less than 10\,pc, to which we recommend measuring the parallax:
J0122+2209/G~34--53 (M4.0\,V, $d$ = 8$\pm$2\,pc) and J2211+4059/1RXS J221124.3+410000 (M5.5\,V, $d$ = 9$\pm$2\,pc).
There are another five stars at 10--15\,pc, four of which are presented here for the first time.
The identification of new relatively bright, low-active, single stars much closer to Earth than the median distance to M-dwarf exoplanet-survey targets ($\gg$13\,pc) is still a matter of interest. 
In summary, this kind of VO color-based search may shed light on the complete identification and characterization of all M dwarfs in the 10\,pc radius sphere centered on the Sun, until the ESA space mission Gaia delivers its final catalog by 2022.

We were genuinely surprised by the discovery of three slow M dwarfs with low surface gravities from weak Na~{\sc i} absorption in our IDS/INT spectra and from VOSA fits to observed multi-wavelength spectral energy distributions.
Besides, one of them, J0422+2439, had a strong H$\alpha$ emission indicative of accretion (the pEW(H$\alpha$)s of the other 26 stars were consistent with the chromospheric activity).
This fact led us to study the X-ray emission of the sample stars in the {\em ROSAT} Bright Source Catalogue.
Of the 11 (40\,\%) positive cross-matches with {\em ROSAT}, 7 were new detections, which suggests that previous all-sky X-ray surveys for low-mass stars have been incomplete.
The coolest star in our sample, J2211+4059 (M5.5\,V), had also the highest $L_X / L_J$ ratio, slightly above those of four other stars, including J0422+2439. 

We assigned membership of J0422+2439 to the Taurus-Auriga star-forming region based not only on low surface gravity and H$\alpha$ and X-ray emissions, but also on coincidence of spatial location, proper motions, and color-magnitude combinations with a large sample of known Taurus-Auriga members.
We also assigned membership of J0435+2523 and J0439+2333, the other two low-gravity stars, in the star-forming region.
The identification of three new intermediate M dwarfs in Taurus-Auriga may help alleviate the reported lack of them, which has made many authors to claim the uniqueness of the initial mass function in Taurus-Auriga. 

We also looked for proper-motion companions to our 27 stars.
We recovered a fragile, wide,  already known pair and reported and characterized a new pair, an M4.5\,V star and an M5.0:\,V companion separated by 6.5\,arcsec ($\sim$110\,AU).

We will continue to search  with VO tools for slow bright nearby M dwarfs, in particular for potential targets for exoplanet hunting.
For that, we will not only plan to conclude the analysis of our CMC14+2MASS data with a new spectroscopic follow-up, but also start a new study with the latest release of the Carlsberg Meridian Catalogue (CMC15\footnote{\tt http://svo2.cab.inta-csic.es/vocats/cmc15/}), which will be more extensive than the CMC14 one.
Extra VO works for the identification of slow bright nearby M dwarfs unnoticed by previous surveys will include massive cross-matches between existing databases relevant for this topic: 2MASS, PPMXL, UCAC4, {\em WISE}, {\em GALEX}, {\em ROSAT}, VISTA, and VST.
These works will pave the way for further `super-massive' correlations when the first {\em Gaia} and {\em EUCLID} data releases are available.\\

\acknowledgments
We gratefully thank F. J. Alonso-Floriano, P.~Cruz-Gamba, A.~Klutsch, and B.~Stelzer for their helpful feedback and data provision.
This publication is based on observations made with the Isaac Newton Telescope operated on the island of La Palma by the Isaac Newton Group in the Spanish Observatorio del Roque de los Muchachos of the Instituto de Astrof\'{\i}sica de Canarias.
This publication has made use of the SIMBAD, VizieR and Aladin, operated at center de Donn\'ees astronomiques de Strasbourg, France, the Washington Double Star Catalog maintained at the U.S. Naval Observatory, and VOSA, a Virtual Observatory tool developed under the Spanish Virtual Observatory project supported from the Spanish MICINN through grant AyA2008-02156. 
MCGO acknowledges the support of a JAE-Doc  CSIC fellowship cofunded with the European Social Fund under the program Junta para la Ampliaci\'on de Estudios.
Financial support was provided by the Spanish Ministerio de Ciencia e Innovaci\'on under grants 
AyA2011-24052 and 
AYA2011-30147-C03-03.

\bibliographystyle{apj}
\bibliography{biblibrary}

%______________________________________________

%__________________________________________________ algo

\end{document}